\documentclass[acmsmall,dvipsnames,screen]{acmart}
\setcopyright{rightsretained}
\acmYear{2024}
\copyrightyear{2024}
\acmSubmissionID{pldi24main-p146-p}
\acmJournal{PACMPL}
\acmVolume{8}
\acmNumber{PLDI}
\acmMonth{6}

\usepackage{booktabs}   
\usepackage{subcaption} 

\setcitestyle{round}

\usepackage{paralist}
\usepackage{listing}
\usepackage[ruled,algosection,noend]{algorithm2e}
\usepackage{multirow}
\usepackage{bussproofs}
\usepackage{varwidth}
\usepackage{graphicx}
\usepackage{listings}
\usepackage{float}
\usepackage{multirow}
\usepackage[noend]{algorithmic}
\usepackage{mathpartir}
\usepackage{url}
\usepackage{hyperref}
\usepackage[capitalise]{cleveref}
\usepackage{xspace}
\usepackage{verbatim}
\usepackage{lipsum}
\usepackage{wrapfig}
\usepackage{ifsym}

\usepackage[inline, shortlabels]{enumitem}
\usepackage{url}
\usepackage{fancyvrb}
\usepackage[figurename=Figure]{caption}
\usepackage{nanoc}
\usepackage[T1]{fontenc}
\usepackage{relsize}
\usepackage{marvosym}
\usepackage{microtype}
\usepackage[cache=true,frozencache]{minted}
\usepackage{adjustbox}
\usepackage{multicol}

\usepackage{xcolor}
\usepackage{tikz}
\usepackage{tcolorbox}
\usepackage{array}
\usepackage{makecell}
\usepackage{pifont}
\newcommand{\cmark}{\ding{51}}%

\usetikzlibrary{arrows}
\usetikzlibrary{shapes.arrows}
\usetikzlibrary{shapes}
\usetikzlibrary{calc}
\usetikzlibrary{positioning}
\usetikzlibrary{fit}
\usetikzlibrary{tikzmark}
\usetikzlibrary{snakes}

\usepackage{tikzsymbols} 

\tikzstyle{arrow}+=[thick,rounded corners=0.5em]
\tikzstyle{every picture}+=[remember picture,baseline]

\usepackage{multirow}
\usepackage{hhline}

\hypersetup{colorlinks,
  linkcolor=ACMDarkBlue,
  citecolor=ACMPurple,
  urlcolor=ACMDarkBlue,
  filecolor=ACMDarkBlue}

\floatname{algorithm}{Algorithm}

\SetAlFnt{\small}
\SetAlCapFnt{\small}
\SetAlCapNameFnt{\small}
\SetAlCapHSkip{0pt}
\IncMargin{-\parindent}

\floatstyle{plaintop}
\restylefloat{table}

\defcitealias{Coq-manual}{Coq}

\makeatletter 
\def\arcr{\@arraycr}
\makeatother

\definecolor{shadecolor}{gray}{1.00}
\definecolor{darkgray}{gray}{0.30}
\definecolor{violet}{rgb}{0.56, 0.0, 1.0}
\definecolor{forestgreen}{rgb}{0.13, 0.55, 0.13}
\definecolor{mygray}{rgb}{0.5,0.5,0.5}

\definecolor[named]{ACMBlue}{cmyk}{1,0.1,0,0.1}
\definecolor[named]{ACMYellow}{cmyk}{0,0.16,1,0}
\definecolor[named]{ACMOrange}{cmyk}{0,0.42,1,0.01}
\definecolor[named]{ACMRed}{cmyk}{0,0.90,0.86,0}
\definecolor[named]{ACMLightBlue}{cmyk}{0.49,0.01,0,0}
\definecolor[named]{ACMGreen}{cmyk}{0.20,0,1,0.19}
\definecolor[named]{ACMPurple}{cmyk}{0.55,1,0,0.15}
\definecolor[named]{ACMDarkBlue}{cmyk}{1,0.58,0,0.21}
\definecolor[named]{PaleGreen}{RGB}{196, 255, 231}
\definecolor[named]{PaleOrange}{RGB}{255, 213, 169}
\definecolor{intnull}{RGB}{213,229,255}

\definecolor{shadecolor}{gray}{1.00}
\definecolor{ddarkgray}{gray}{0.85}
\definecolor{darkgray}{gray}{0.30}
\definecolor{light-gray}{gray}{0.91}

\definecolor{ipurple}{HTML}{E5D8F3}
\definecolor{igreen}{HTML}{C2F4CF}
\definecolor{ired}{HTML}{AE0428}
\definecolor{iblue}{HTML}{3E74D1}
\definecolor{ilgray}{HTML}{E3E3E3}
\definecolor{idgray}{HTML}{747474}

\newcommand{\angled}[1]{\left\langle{#1}\right\rangle}

\newcommand{\ie}{\emph{i.e.}\xspace}

\newcommand{\eg}{\emph{e.g.}\xspace}

\newcommand{\aka}{\textit{a.k.a.}\xspace}
\newcommand{\cf}{\textit{cf.}\xspace}

\newcommand{\wrt}{\emph{w.r.t.}\xspace}

\newcommand{\eqdef}{\triangleq}

\newcommand{\tname}[1]{\textsc{#1}\xspace}

\newcommand{\logicfull}{Logic for Graceful Tensor Manipulation\xspace}
\newcommand{\logic}{LGTM\xspace}

\newcommand{\Local}{{\tt local}\xspace}

\renewcommand{\wp}[2]{{\mathsf{wp}}~\left[ #1 \right]~{\color{grown}\left\{ #2 \right\}}}
\newcommand{\wpf}[3]{{\mathsf{wp}}~\left[ #1 \right]~{\color{grown}\left\{#2~\vert~#3 \right\}}}
\newcommand{\local}[2]{\Local(#1,#2)}

\newcommand{\Loc}{\mathsf{Loc}}
\newcommand{\Val}{\mathsf{Val}}

\definecolor{pblue}{rgb}{0.13,0.13,1}
\definecolor{pgreen}{rgb}{0,0.5,0}
\definecolor{pred}{rgb}{0.9,0,0}
\definecolor{pgrey}{rgb}{0.46,0.45,0.48}

\definecolor{ckeyword}{HTML}{7F0055}
\definecolor{ccomment}{HTML}{3F7F5F}
\definecolor{cnumber}{HTML}{2A0099}

\newcommand{\code}[1]{\text{\mintinline[fontsize=\small]{python}{#1}}}
\newcommand{\fcode}[1]{\text{\mintinline[fontsize=\scriptsize]{python}{#1}}}

\newcommand{\x}{\times}

\newcommand{\set}[1]{\left\{{#1}\right\}}
\newcommand{\many}[1]{\overline{#1}}

\makeatletter
\protected\def\ccell#1#{%
  \kern-\fboxsep
  \@ccell{#1}%
}
\def\@ccell#1#2#3{%
  \colorbox#1{#2}{#3}%
  \kern-\fboxsep
}
\makeatother

\newcommand{\pts}{\mapsto}
\newcommand{\True}{\mathsf{true}}
\newcommand{\False}{\mathsf{false}}

\newcommand{\sep}{\ast}

\newcommand{\rulename}[1]{\textsc{#1}}

\newcommand{\kw}[1]{{\tt{\small{\color{forestgreen}{#1}}}}\xspace}

\newcommand{\qmark}{{\tt{\small{?}}}}

\newcommand{\mcode}[1]{{\text{\code{#1}}}}

\newcommand{\hide}[1]{}
\newcommand{\bientails}{\dashv\vdash}

\makeatletter
\providecommand*{\cupdot}{%
  \mathbin{%
    \mathpalette\@cupdot{}%
  }%
}
\newcommand*{\@cupdot}[2]{%
  \ooalign{%
    $\m@th#1\cup$\cr
    \hidewidth$\m@th#1\cdot$\hidewidth
  }%
}
\makeatother

\DeclareMathSymbol{\synth}{\mathrel}{symbolsC}{123}

\usetikzlibrary{decorations.pathmorphing}

\tikzset{snake it/.style={decorate, decoration=snake}}

\def\signed #1{{\leavevmode\unskip\nobreak\hfil\penalty50\hskip1em
  \hbox{}\nobreak\hfill #1%
  \parfillskip=0pt \finalhyphendemerits=0 \endgraf}}

\newsavebox\mybox

\providecommand{\str}{ \ * \ }
\providecommand{\sep}{\str}

\newcommand{\ith}{$^{\text{th}}$\xspace}

\providecommand{\forl}[4]{\kw{for}~#1~\kw{in~range}(#2,#3) \ \left\{ #4 \right\}}

\providecommand{\ifc}[2]{\kw{if}~#1~\kw{then}~#2}
\providecommand{\whilel}[2]{\kw{while}\left(#1\right)\left\{ #2 \right\}}

\providecommand{\Pm}{P_{\code{mc}}(\cdot)}
\providecommand{\Pv}{P_{\code{vc}}(\cdot)}

\definecolor{grown}{rgb}{0.6, 0, 0.6}
\definecolor{blue}{rgb}{0, 0, 0.9}

\providecommand{\hsll}[4]{
    {\color{grown}\left\{\begin{array}{ll} #1 \end{array} \right\}}
    \left[
    #2  
    \right]
    {\color{grown}\left\{\!\!\left.\begin{array}{c} #3 \end{array}\!\right|\! \begin{array}{ll} #4 \end{array} \right\}}
}

\newcommand{\hspec}[1]{{\color{grown}\set{#1}}}
\newcommand{\hmid}[1]{\left[#1\right]}
\newcommand{\pp}{\mathcal{P}}

\providecommand{\hsllazy}[4]{
    {\color{grown}\left\{
      \begin{array}{ll} 
        \!\! \!\!
        \begin{array}{l@{\ }c}
          #1
        \end{array}
        \!\! \!\! \!\! 
      \end{array} 
      \right\}}
    \left[
      {\!\!
        \begin{array}{c@{\ }c@{\ }l}
          #2
        \end{array}
        \!\!}
    \right]
    {\color{grown}\left\{\!\!\left.\begin{array}{l} #3 \end{array}\!\right|\! \begin{array}{ll}\!\!  #4 \end{array} \right\}}
}

\providecommand{\hslc}[4]{{\color{grown}\left\{ #1 \right\}}\left[ #2 \right] {\color{grown}\left\{ #3~\left|~#4\right. \right\}} }
\providecommand{\hslcc}[3]{{\color{grown}\left\{ #1 \right\}}\left[ #2 \right] {\color{grown}\left\{#3\right\}} }

\providecommand{\arr}[2]{\mathtt{arr}(\mathtt{#1},\
  #2)}
 \providecommand{\arrl}[2]{\mathtt{arr}(#1,\ #2)}

\providecommand{\rule}[3]{
  \begin{prooftree}
    \AXC{\visible<2->{#1}}
    \RightLabel{\visible<2->{#2}}
    \UIC{#3}
  \end{prooftree}
}

\newcommand{\bigstr}{\mathop{\scalebox{2.5}{\raisebox{-0.2ex}{$\ast$}}}}
\providecommand{\pure}[1]{\lceil #1 \rceil}
\newcommand{\tinds}[1]{\mathtt{#1}_{\mathtt{ind}}}
\newcommand{\tvals}[1]{\mathtt{#1}_{\mathtt{val}}}
\newcommand{\tidxs}[1]{\mathtt{#1}_{\mathtt{idx}}}
\newcommand{\inds}[1]{{#1}_{\mathit{ind}}}
\newcommand{\vals}[1]{{#1}_{\mathit{val}}}
\newcommand{\idxs}[1]{{#1}_{\mathit{idx}}}

\newcommand{\arrs}{\mathit{Arrs}}
\newcommand{\Zero}{\mathbf{0}}
\newcommand{\Ispmspv}{I_{\kw{for}}\xspace}
\newcommand{\Ispvspv}{I_{\kw{while}}\xspace}

\newcommand{\Ans}{{\mcode{ans}}}

\newcommand{\ind}{{\mathit{ind}}}

\newcommand{\Iyy}{\mathit{i_y}}
\newcommand{\Ixx}{\mathit{i_x}}

\definecolor{shadecolor}{gray}{1.00}
\definecolor{darkgray}{gray}{0.30}
\definecolor{violet}{rgb}{0.56, 0.0, 1.0}
\definecolor{forestgreen}{rgb}{0.13, 0.55, 0.13}
\definecolor{mygray}{rgb}{0.5,0.5,0.5}
\definecolor{mahogany}{RGB}{192, 64, 0}

\lstdefinelanguage{Coq} {
mathescape=true,						
texcl=false,
morekeywords=[1]{
  Add,
  All,
  Arguments,
  Axiom,
  Bind,
  Canonical,
  Check,
  Close,
  CoFixpoint,
  CoInductive,
  Coercion,
  Contextual,
  Corollary,
  Defined,
  Definition,
  Delimit,
  End,
  Example,
  Export,
  Fact,
  Fixpoint,
  Goal,
  Graph,
  Hint,
  Hypotheses,
  Hypothesis,
  Implicit,
  Implicits,
  Import,
  Inductive,
  Lemma,
  Let,
  Local,
  Locate,
  Ltac,
  Maximal
  Module,
  Morphism,
  Next,
  Notation,
  Obligation,
  Open,
  Parameter,
  Parameters,
  Prenex,
  Print,
  Printing,
  Program,
  Projections,
  Proof,
  Proposition,
  Qed,
  Record,
  Relation,
  Remark,
  Require,
  Reserved,
  Resolve,
  Rewrite,
  Save,
  Scope,
  Search,
  Section,
  Show,
  Strict,
  Structure,
  Tactic,
  Theorem,
  Unset,
  Variable,
  Variables,
  View,
  inside,
  outside
},
morekeywords=[2]{
  as,
  cofix,
  false,
  else,
  end,
  exists,
  exists2,
  fix,
  for,
  forall,
  fun,
  if,
  in,
  is,
  let,
  match,
  nosimpl,
  of,
  return,
  struct,
  then,
  true,
  vfun,
  with
},
morekeywords=[3]{Type, Prop, Set, True, False},
morekeywords=[4]{
  ptr, seq, nat, unit, heap,
  after,
  apply,
  assert,
  auto,
  bool_congr,
  case,
  change,
  clear,
  compute,
  congr,
  constructor,
  cut,
  cutrewrite,
  destruct,
  elim,
  field,
  fold,
  generalize,
  have,
  hhauto,
  heval, 
  hnf,
  induction,
  injection,
  intro,
  intros,
  intuition,
  inversion,
  left,
  loss,
  move,
  nat_congr,
  nat_norm,
  pattern,
  pose,
  refine,
  rename,
  replace,
  revert,
  rewrite,
  right,
  ring,
  set,
  simpl,
  split,
  suff,
  suffices,
  symmetry,
  transitivity,
  trivial,
  unfold,
  unlock,
  using,
  without,
  wlog,
  autorewrite
},        
morekeywords=[5]{
  assumption,
  by,
  contradiction,
  done,
  exact,
  lia,
  gappa,
  omega,
  reflexivity,
  romega,
  solve,
  tauto,
  discriminate,
  unsat
},
morecomment=[s]{(*}{*)},
morekeywords=[6]{do, first, try, idtac},
showstringspaces=false,
morestring=[b]",
keepspaces=true, 
tabsize=3,							
extendedchars=true,  		 		
sensitive=true, 
breaklines=false,
basicstyle=\footnotesize\ttfamily,
captionpos=b,							
columns=[l]fullflexible,
identifierstyle={\color{black}},
keywordstyle=[1]{\color{violet}},
keywordstyle=[2]{\color{forestgreen}\bfseries},
keywordstyle=[3]{\color{forestgreen}\bfseries},
keywordstyle=[4]{\color{blue}},
keywordstyle=[5]{\color{red}},
keywordstyle=[6]{\color{violet}},
stringstyle=\it\ttfamily\color{mahogany},
commentstyle=\it\ttfamily\color{Bittersweet},
numberstyle=\tiny\color{mygray},
literate={\\/}{{$\vee$}}1
         {/\\}{{$\wedge$~}}1
         {:->}{{$\mapsto~$\!}}1
         {ent}{{$\vdash~$}}1
         {\\in}{{$\in~$}}1
         {++}{{$+\!+\!~$}}1
         {.+}{{$\!\!+$}}1
         {-->}{{$\rightarrow$}}1
         {forall}{{$\forall~$}}1
         {not}{{$\neg~$}}1
         {exists}{{$\exists~$}}1
         {sep}{{$\sep~$}}1
        {=>}{{$\Rightarrow~$}}1
         {\\+}{{$\!\join\!~$}}1
         {\\x}{{$\mathtt{x}~$}}1
         {\\cap}{{$\mathtt{\cap}~$}}1
         {\\phi}{{$\mathtt{\phi}~$}}1
         {midx}{{$\mathtt{m_{idx}}~$}}1
         {mind}{{$\mathtt{m_{ind}}~$}}1
         {mval}{{$\mathtt{m_{val}}~$}}1
         {xind}{{$\mathtt{x_{ind}}~$}}1
         {xval}{{$\mathtt{x_{val}}~$}}1
         {\\ss}{{$\S~$}}1
         {inv}{{$\mathtt{I_{\color{forestgreen}for}}~$}}1
         {capsigma}{{$\Sigma~$}}1
         {lceil}{{$\lceil~$}}1
         {rceil}{{$\rceil~$}}1
         {alpha}{{$\alpha~$}}1
         {beta}{{$\beta~$}}1
}

\lstdefinestyle{Coq}{language=Coq}

\usetikzlibrary{matrix}
\usetikzlibrary{calc}

\setlist[itemize]{leftmargin=*}
\setlist[enumerate]{leftmargin=*}

\allowdisplaybreaks
\begin{document}

\newcommand{\mytitle}{Mechanised Hypersafety Proofs about Structured Data}
\newcommand{\runningtitle}{\mytitle}

\setlength\floatsep{1.25\baselineskip plus 3pt minus 2pt}
\setlength\textfloatsep{1.25\baselineskip plus 3pt minus 2pt}
\setlength\intextsep{1.25\baselineskip plus 3pt minus 2 pt}

\title[\runningtitle]{\mytitle}
\subtitle{Extended Version}

\author{Vladimir Gladshtein}
\affiliation{%
  \institution{National University of Singapore}
    \country{Singapore}
}
\email{vgladsht@comp.nus.edu.sg}
\orcid{0000-0001-9233-3133}

\author{Qiyuan Zhao}
\affiliation{%
  \institution{National University of Singapore}
    \country{Singapore}
}
\email{qiyuanz@comp.nus.edu.sg}
\orcid{0000-0002-1017-1562}

\author{Willow Ahrens}
\affiliation{%
  \institution{Massachusetts Institute of Technology}
    \country{USA}
}
\email{willow@csail.mit.edu}
\orcid{0000-0002-4963-0869}

\author{Saman Amarasinghe}
\affiliation{%
  \institution{Massachusetts Institute of Technology}
    \country{USA}
}
\email{saman@csail.mit.edu}
\orcid{0000-0002-7231-7643}

\author{Ilya Sergey}
\affiliation{%
  \institution{National University of Singapore}
    \country{Singapore}
}
\email{ilya@nus.edu.sg}
\orcid{0000-0003-4250-5392}

\begin{abstract}

Arrays are a fundamental abstraction to represent collections of data.
It is often possible to exploit \emph{structural properties} of the
data stored in an array (\eg, repetition or sparsity) to develop a
specialised representation optimised for space efficiency.
Formally reasoning about correctness of manipulations with such
structured data is challenging, as they are often composed of multiple
loops with non-trivial invariants.
%


In this work, we observe that specifications for structured data
manipulations can be phrased as \emph{hypersafety} properties, \ie,
predicates that relate traces of $k$~programs.
%
%
To turn this observation into an effective verification methodology,
we developed the \logicfull (\logic), a new Hoare-style relational
separation logic for specifying and verifying computations over
structured data.
The key enabling idea of \logic is that of \emph{parametrised}
hypersafety specifications that allow the number~$k$ of the program
components to depend on the \emph{program variables}.
%
%
We implemented \logic as a foundational embedding into Coq,
mechanising its rules, meta-theory, and the proof of soundness. 
Furthermore, we developed a library of domain-specific tactics that
automate computer-aided hypersafety reasoning, resulting in
pleasantly short proof scripts that enjoy a high degree of reuse.
We argue for the effectiveness of relational reasoning about
structured data in \logic by specifying and mechanically proving
correctness of 13 case studies including computations on compressed
arrays and efficient operations over multiple kinds of sparse tensors.

\end{abstract}



\maketitle


\section{Introduction}
\label{sec:intro}


Arrays are one of the most basic yet most powerful abstractions in
computer science, capable of representing virtually any kind of data.
The most common way to encode data is via a \emph{dense} array that
stores data in memory \emph{contiguously}, providing a simple
interface to access it via iteration.
Representing data via dense arrays is often not optimal in terms of
space efficiency.
For instance, a \emph{sparse tensor} (\ie, an $n$-dimensional matrix
with most of its elements being zeros) can be encoded compactly by
storing only its non-zero elements. Without this space optimisation,
manipulations with tensors that contain real-world datasets becomes
impossible due to the immense amount of storage that would be required
by their dense representations.


Sparsity of a dense array is just one of the many structural
properties of the array-stored data; others include \emph{symmetry}
(\eg, an array-encoded matrix is equal to its transpose),
\emph{repetition} (data contains segments of repeating elements), or
length-\emph{irregularity} (\aka ragged arrays).
Realising such data properties allows one to choose a specialised
\emph{format} for representing \emph{structured data}, leading to
reduced size of the required storage and improved performance by
avoiding redundant computations (\eg, skipping through all zeros when
performing summation over a sparse tensor).

%
%
Unsurprisingly, computations that involve structured data are
typically much more intricate than simply iterating through a dense
array, as they should take the conventions imposed by the data's
format into account.
The problem of automatically generating code, which is specialised for
particular representations of structured data it manipulates with, has
received a lot of attention in recent years.
Some of the most prominent works in this direction include the
influential TACO framework for code generation~\cite{KjolstadKCLA17}
and the \tname{Finch} compiler~\cite{AhrensDKA23}.

%
\paragraph{Problem statement}

Given the intricacy of computations on structured data, it is natural
to wonder: \emph{Do they produce the same results as their
  non-specialised counterparts on dense data representations?}
This question suggests a verification challenge that we focus on in
this work.

%
Existing efforts on verifying computations with structured data either
provide high-level \emph{domain-specific languages} (DSL) for encoding
data formats as well as format-aware computations~\cite{ArnoldHKBS10},
or
encode data formats as \emph{logical predicates} in a language of a
solver used to verify refinement \wrt operations on the dense
representations~\cite{DyerAB19}.
By offering a limited set of predefined abstractions, these DSL- or
predicate-based approaches do not allow for verifying formats, whose
encoding does not fit into their formalisms.
None of these approaches come with a formal soundness proof regarding
the executable code they produce.
A recent work by~\citet{LAproof} presented a mechanised correctness
proof of a particular sparse matrix computation in a general-purpose
separation logic embedded into Coq~\cite{Appel:ESOP11}. That
verification effort relied on a tailored encoding of the chosen data
format and required several format-specific lemmas, making it
difficult to reuse any of its components for verifying programs over
different formats.

The goal of this work is to provide a versatile framework for formally
specifying and mechanically verifying diverse computations with
structured data.
We aim for an approach that 
(a)~allows for naturally-looking and easy-to-state specifications, 
(b)~can accommodate arbitrary programmatically-defined data formats,
and
(c)~enables a high degree of proof reuse and automation.

In the rest of the paper, we argue that \emph{relational program
  logics} are the right tool for this job.

\paragraph{Specifying manipulations with structured data}
\begin{wrapfigure}[11]{r}{0.43\textwidth}
  \vspace{-15pt}
  \begin{minted}[numbers=left, xleftmargin=2.5ex, numbersep=6pt, escapeinside=@@, fontsize=\footnotesize]{python3}
  def m_get (mc, i, j) -> val: 
    # matrix format; body omitted

  def v_get (vc, i) -> val:    
    # vector format; body omitted

  def mv_prod(mc, vc) -> val[]:
    let ans = malloc(vc.size)
    for mp in partition(mc):
      inner_prod (mp, vc, ans)
  \end{minted}
  \setlength{\abovecaptionskip}{10pt}  
  \caption{Structured matrix/vector multiplication.}
  \label{fig:sigs}
\end{wrapfigure} 
Let us immediately dive into computations with structured data.
\autoref{fig:sigs} shows a high-level outline of one such computations
in Python-style pseudocode. The function \code{mv_prod} multiplies a
compressed matrix, compactly encoded by a data structure \code{mc}, by
a compressed vector, encoded via \code{vc}.\footnote{Compressed
  tensors, such as \fcode{mc} and \fcode{vc} from \autoref{fig:sigs},
  are typically encoded as tuples of 1-dimensional arrays with their
  sizes, but might be also represented as linked data
  structures~\cite{ChouA22}.}
The two random access functions, \code{m_get} and \code{v_get}, whose
signatures are given, \emph{specify} the formats used to compress the
matrix and the vector, correspondingly.
For instance, given a compressed matrix \code{mc} and two indices (row
and column position), \code{m_get} returns a value (of some type
\code{val}) stored in the ``algebraic'' (\ie, dense) version of the
matrix at those positions, without fully reconstructing the dense
representation.

An efficient implementation \code{mv_prod} of a matrix/vector product,
therefore, operates \emph{directly} on the compressed representations
\code{mc} and \code{vc} provided as its arguments. Such computations
are typically structured by partitioning the representation of the
matrix into separate chunks in some format-specific way, and storing
the results of submatrix/vector products into the (dense) result
vector \code{ans}.
Functions of this sort, operating on sparse tensors, compressed via
multiple different formats, are standard components of popular
libraries for tensor computations, such as \tname{SciPy} (Python),
\tname{CsMatBase} (Rust), and \tname{Eigen} (C++).
While at the time of this writing most of such function
implementations are hand-crafted, they can be also automatically
generated by domain-specific compilers, such as
\tname{Finch}~\cite{AhrensDKA23}, provided the corresponding
computation on dense matrices/vectors and the programmatic format
specifications such as \code{m_get} and \code{v_get}.

Importantly, the programmatic format definitions \code{m_get} and
\code{v_get} are \emph{not} used directly within the implementation of
\code{mv_prod}. Instead, they should be considered as
\emph{specifications} of the compressed data that \code{mv_prod}
manipulates with.
Given such specifications, modern domain-specific compilers are
capable of producing efficient format-specific implementation from
\emph{reference} implementations of the operations phrased in terms of
dense representations~\cite{AhrensDKA23,SakkaS017}.

How do we specify correctness of \code{mv_prod}, asserting that its
result is the same as a multiplying an \emph{uncompressed} version of
\code{mc} to that of \code{vc}?
The key observation we make in this work is that such correctness
statements can be stated naturally and concisely as Hoare-style
triples that \emph{relate} executions of the programmatic format
specifications, such as \code{m_get} and \code{v_get}, to that of the
operation in question.
A common way to capture it formally is via a \emph{hypersafety}
judgment---a logical statement that relates results of executions of
\emph{multiple} possibly different computations.
Hypersafety properties can be ascribed to a product of
\emph{independently} run programs in a form of (Hoare-style)
\emph{hyper-triples}, whose pre/postconditions constrain collections
of pre/post-states of the programs in question, as well as their
results.
Therefore, a correctness specification for \code{mv_prod} can be
phrased as the following hyper-triple relating its result to those of
\code{m_get} and \code{v_get}:
\begin{equation}
  \label{eq:mvprod}
  {\small{
  \hspec{\!
  \begin{array}{l}
    \Pm \str \Pv
  \end{array}
  \!\!}    
  \hmid{\!
    \begin{array}{c@{}c@{\ \ \ }l}
    1 &:& \mathtt{mv\_prod(mc, vc)} 
    \\ [2pt]
    \angled{2, i, j}_{\!\!\!\!\tiny
      \begin{array}{l} i \in [0, M) \\ j \in [0,N) \end{array}} &:&  \mathtt{m\_get(mc,\ } i,\ j)
    \\ [2pt]
    \angled{3, j}_{j \in [0, N)} &:& \mathtt{v\_get(vc,\ } j)
    \end{array}
  \!}
  \hspec{\!\!\!
  \left.
  \begin{array}{r}
  a \\[2pt] \many{m} \\[2pt] \many{v}
  \end{array}
  \right|
  \begin{array}{l}
  \exists s, \arrl{a}{s} \str
  \\[2pt]
  \pure{\forall i, s[i] = \sum_j \many{m}(i,j)\many{v}(j)} 
  \end{array}
  \!\!\!}
  }}
\end{equation}

The hypersafety specification~\eqref{eq:mvprod} relates the execution
of $1 + M \times N + N$ programs, with $M$ and $N$ being the number of
the rows and the columns in the matrix.
For instance, the component
$\angled{2,i,j}_{i \in [0,M),\, j \in [0,N)} : \mcode{m_get}(m, i, j)$
of the triple represents a \emph{collection} of independent runs of
\code{m_get} \emph{indexed} by tuples~$\angled{2, i, j}$, with~$i$
ranging from~$0$ to~$M - 1$ and~$j$ from~$0$ to~$N - 1$.
In such cases, we will refer to the set like
$\set{\angled{2,i,j} \mid {i \in [0,M),\, j \in [0,N)}}$ as
\emph{index sets}.
The results of those $M \times N$ programs are bound to a sequence
(hyper-)variable $\many{m}$ of size $M \times N$, whose individual
components are retrieved in the postcondition as $\many{m}(i,j)$.
The same intuition applies to the last $N$ program components, whose
results are bound to~$\many{v}$.
The postcondition of the triple, thus, ensures that the result of the
first program component $a$, is a base pointer of an array storing a
mathematical sequence $s$, which is element-wise equal to the
dot-product of the given matrix and vector.

The precondition features two Separation Logic-style predicates,
$P_{\code{mc}}$ and $P_{\code{vc}}$, that define the memory shape of
the corresponding compressed matrix/vector representations (\ie, state
that they are encoded in memory by array tuples). 
The definitions of those predicates are quite unremarkable, as they do
not capture any relations between elements of a compressed encoding
and its dense counterpart: that task is left to \code{m_get} and
\code{v_get}.
The only noteworthy component of the precondition is the $(\cdot)$
notation, which indicates \emph{distinct copies} of the same symbolic
state across multiple state components indexed by elements of
$\set{1}\cup \set{\angled{2,i,j} \mid \ldots}\cup\set{\angled{3,j}
  \mid \ldots }$, thus, supplying a valid (\ie, safe to execute from)
precondition to each one of the $1 + M \times N + N$ programs in the
hyper-triple.


\paragraph{Key idea}

The specification~\eqref{eq:mvprod} of \code{mv_prod} owes its brevity
to the idea of \emph{parametrised} hyper-triples.
Specifically, parametrisation means that the index sets used in
program components can be expressed as \emph{functions of the input
  state and variables} of the programs in question.
To wit, the index sets in~\eqref{eq:mvprod} are determined by the
variables $M$ and $N$, which correspond to the ``logical'' dimensions
of the involved matrix, and are constrained by the predicate
$P_{\code{mc}}$.\footnote{In our example, we made those variables
  explicit in the specification, omitting the full signature of
  $P_{\texttt{mc}}$ for brevity.}
The benefits of parametrisation are not limited to relatively concise
specifications.
While very simple conceptually, the ability to vary the number of the
components in a hyper-triple depending on logical- and program-level
variables is surprisingly useful for verifying properties of iterative
computations on non-overlapping data, such as formats for compression
or sparsity.
As we will show in \autoref{sec:overview}, the idea of parametrised
hyper-triples makes it almost straightforward to decompose proofs of
(stated relationally) correctness of compressed structure-manipulating
programs featuring multiple loops.
%

\paragraph{Towards practical reasoning about parametrised hypersafety}
So-called \emph{relational} program logics are a well-studied
formalism to specify and reason symbolically about hypersafety
properties of programs, such as information-flow security and
correctness of optimisations~\cite{Benton04,BartheCK11,Yang07,
  CarbinKMR12, NanevskiBG11, BartheKOB12}.
%
%
%
Alas, the vast majority of existing relational logics are limited to
specifying and proving 2-properties only, while our approach,
exemplified by the spec~\eqref{eq:mvprod}, requires proofs about
arbitrary-arity parametrised specifications.

Two relatively new logics, CHL by \citet{SousaD16} and LHC by
\citet{DOsualdoFD22}, provide proof rules for arbitrary-arity
hypersafety triples.
Unfortunately, CHL does not feature the necessary proof composition
principles: it fixes the arity of a hyper-triple for the entire proof,
making it impossible to employ specifications about subsets of the
program components (or Hoare-style proofs for individual programs) in
the context of a larger proof.
LHC, on the other hand, allows for combining specifications of
arbitrary \emph{constant} arity in a single proof, but lacks rules for
working with hyper-triples, whose arity depends on \emph{program-level
  variables}.
This limitation of LHC turns out to be a show-stopper for constructing
proofs about loops \emph{within the logic}, requiring one to resort to
proofs \emph{in terms of semantics}.
Furthermore, LHC is not a Separation Logic and, therefore, provides no
support for modular reasoning about arrays and heap-based indirection.
Finally, LHC only exists on paper: it is not implemented as a tool, and
its soundness has not been mechanised.

In our quest of building a framework for machine-assisted verification
of structured data manipulations, which is both foundational (\ie,
proven sound from first principles) and practical (\ie, requires low
annotation overhead), we are taking forward the ideas for
compositional hypersafety reasoning pioneered by
LHC~\cite{DOsualdoFD22}, enhancing them with three new aspects:

\begin{enumerate} 
\item adding new rules for decomposing proofs about \emph{arbitrary
    arity-parametric} hyper-tripes;
\item providing specialised \emph{proof rules for loops} that take
  advantage of hyper-triple parametrisation;
\item introducing support for \emph{state-local} hypersafety proofs in the style
  of Separation Logic.
\end{enumerate}

\noindent
The result is the novel logical framework for Hoare-style hypersafety
proofs, dubbed \emph{\logicfull} (\logic).
We implemented \logic as a Separation Logic~\cite{Reynolds:LICS02}
embedded into the Coq proof assistant.
The shallow nature of the embedding allowed us to use Coq's
higher-order logic capabilities for implementing rules for handling
parametrised triples in the form of ordinary Coq lemmas.
Furthermore, the careful design of rules for hyper-triple
decomposition and reasoning about loops made it possible to engineer
effective proof automation resulting in very short and conceptually
simple machine-assisted verification proofs.

\vspace{-2pt}

\paragraph{Contributions and outline}

\vspace{-2pt}

To summarise, in this work we make the following contributions.

\begin{itemize}

\item Our first conceptual contribution is an observation that
  specification and verification of manipulations with structured data
  can be phrased in terms of reasoning about hypersafety properties.

\item Our second conceptual contribution is an idea of a
  \emph{parametrised} hyper-triple, whose arity depends on attributes
  of the data under manipulation.
  We show that parametrisation enables hypersafety proofs based on the
  structure of the \emph{data} rather than the structure of the
  \emph{program} (\autoref{sec:overview}).

\item Our main theoretical contribution is the \logicfull (\logic)---a
  Hoare-style relational separation logic for hypersafety properties
  that provides effective reasoning principles about
  array-manipulating programs with loops (\autoref{sec:logic}).
  We mechanised the meta-theory, the rules, and the soundness proof of
  \logic in the Coq proof assistant.


\item Our practical contribution is a set of Coq-based automation
  techniques for mechanised proofs in \logic that allow for short and
  intellectually manageable hypersafety proofs (\autoref{sec:mech}).
  We have evaluated our implementation of \logic by mechanising proofs
  for 13~characteristic case studies that implement computations with
  sparse and compressed tensors in various formats
  (\autoref{sec:eval}).
  We show that, in comparison with related foundational efforts,
  mechanised proofs in \logic are about 10 times shorter than those
  done in a general-purpose non-relational Separation Logic.

\end{itemize}


%


\providecommand{\ind}{\mathit{ind}}

\section{A Tour of \logic}
\label{sec:overview}

\EnableBpAbbreviations

In this section, we showcase the reasoning principles of \logic as a
logic formalism, postponing the demonstration of its verification
capabilities as a tool until \autoref{sec:mech}~and~\ref{sec:eval}.
To do so, we specify and verify a program that computes a product of a
compressed sparse matrix and a sparse vector, represented via the
Unordered Compressed Sparse Row (uCSR) format and a Sparse Vector
(SV), correspondingly, storing the result into an uncompressed vector
represented by an array.

\begin{figure}[t]
\begin{tabular}{cc}
\begin{minipage}{0.47\linewidth}
\begin{subfigure}{1.0\linewidth}
\begin{tabular}{c}
\includegraphics[width=0.55\linewidth]{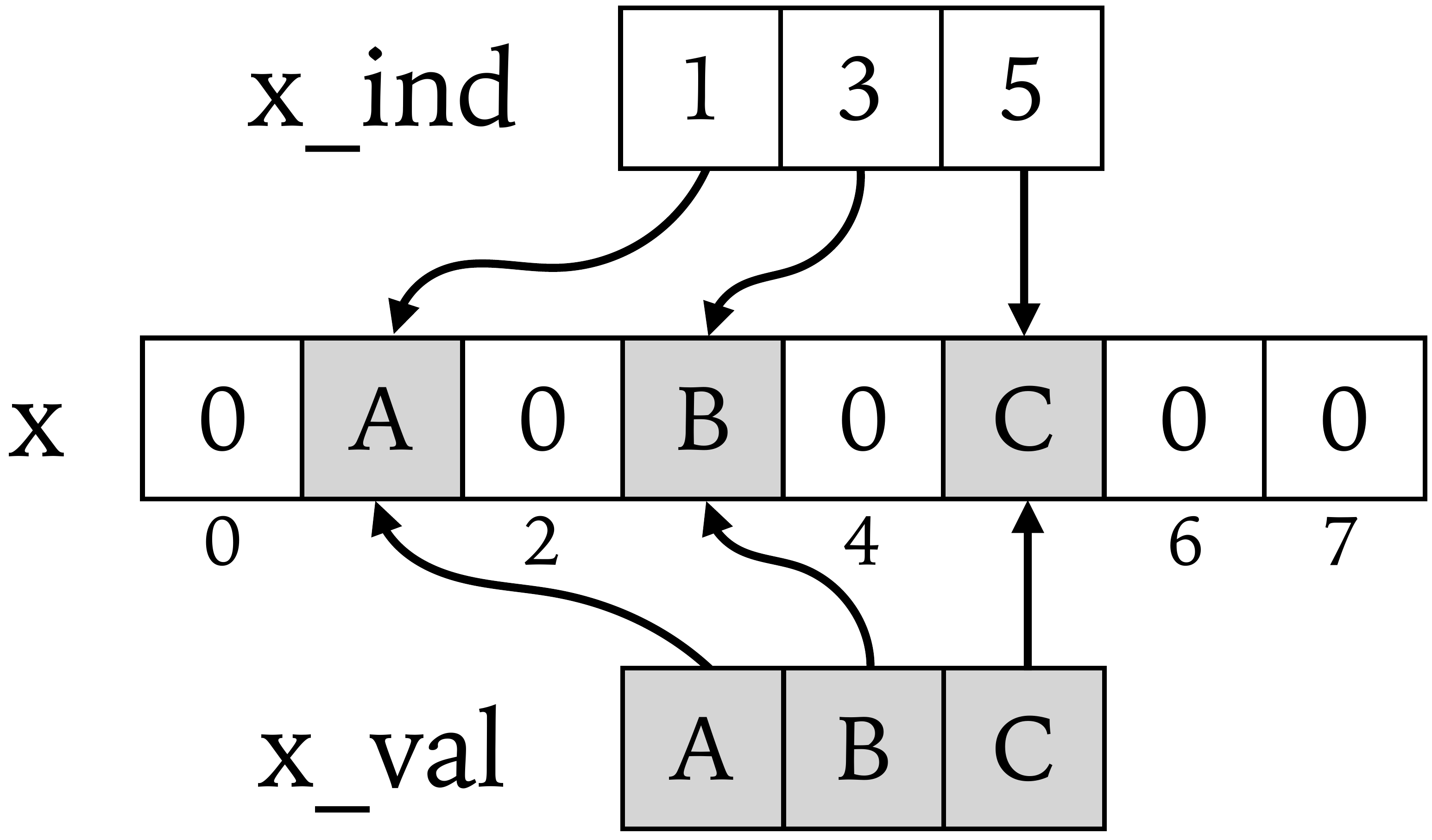}    
\\
\vspace{10pt}
\includegraphics[width=0.90\linewidth]{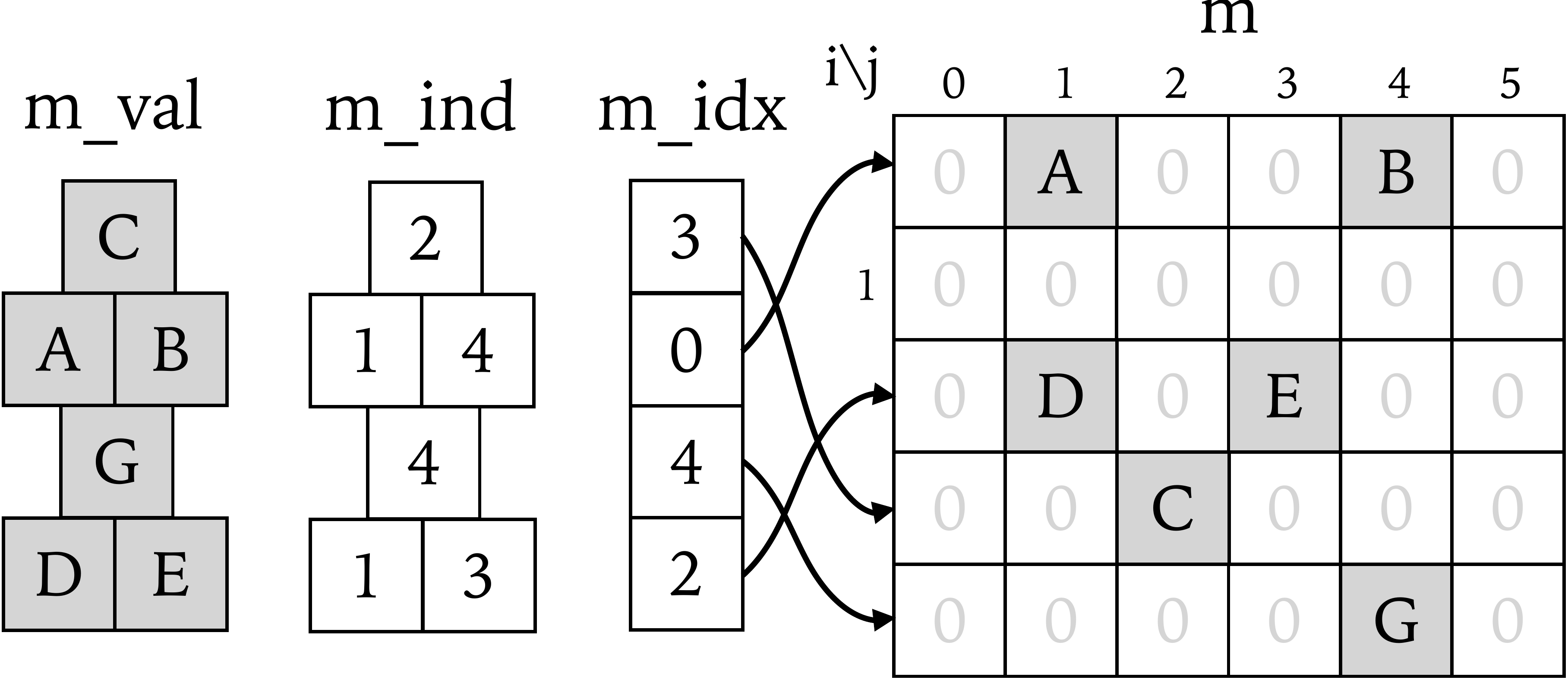}      
\end{tabular}
\vspace{-12pt}
\caption{SV and uCSR, pictographically}
\label{fig:ucsr-pics}                  
\end{subfigure}
\end{minipage}
&
\begin{tabular}{c}
\begin{subfigure}{0.47\textwidth}
\begin{minted}[numbers=left, xleftmargin=2.0ex, numbersep=6pt, escapeinside=@@, fontsize=\footnotesize]{python3}
def sv_get(x_ind, x_val, i):
  ind = index(x_ind, i) 
  if (ind != -1):
    return x_val[ind]
  else: 
    return 0
    \end{minted}
    \caption{Sparse Vector (SV) format}
    \label{fig:sv}    
\end{subfigure}    
\\
\begin{subfigure}{0.47\textwidth}
  \vspace{2mm}
\begin{minted}[numbers=left, xleftmargin=2.0ex, numbersep=6pt, escapeinside=@@, fontsize=\footnotesize]{python3}
def ucsr_get(m_idx, m_ind, m_val, i, j):
  ind = index(m_idx, i)
  if (ind != -1):
    return sv_get(m_ind[ind], m_val[ind], j)
  else: 
    return 0
\end{minted}
\caption{Unordered Compressed Sparse Row (uCSR)}
\label{fig:ucsr}  
\end{subfigure}    
\end{tabular}
\end{tabular}
\caption{Sparse Vector and Unordered Compressed Sparse Row formats}
\label{fig:ucsr-formats}  
\end{figure}

\autoref{fig:ucsr-pics} graphically depicts the data layout in both SV
and uCSR formats.
The SV format is a standard representation for sparse vectors. 
It achieves great rates of compression for sparse (\ie, mostly-zero)
vectors by storing only their \emph{non-zero} values in an array
\code{x_val} together with their indices at the same positions in an
array \code{x_ind}.
The programmatic encoding of SV is provided by the \code{sv_get}
function shown in~\autoref{fig:sv}. 
This function ``decompresses'' a sparse vector encoded by the pair of
arrays (\code{x_val}, \code{x_ind}) by retrieving the \code{i}\ith
value of the dense representation \code{x} from its compressed
representation via \code{x_ind} and \code{x_val}. 
It does so by first finding the index of the element corresponding to
\code{i} in the array \code{x_ind}, calling the \code{index} function
(line~2). 
The implementation of \code{index} simply performs a linear scan
through the contents of \code{x_ind}, returning a position of its
second argument, if it's present, and -1 otherwise (remember,
\code{sv_get} only serves as a format specification, hence its
efficiency is not important).
Depending on the search result stored into \code{ind}, the rest of
\code{sv_get}'s implementation returns either the corresponding
\code{x_val} element, or zero (lines~3-6).

The uCSR format for sparse matrices is a modification of the popular
Compressed Sparse Row~(CSR) format, designed to be more efficient by
enabling adaptive tiling~\cite{ChangwanRIS19}.
uCSR adopts a compression principle similar to that of SV: rather than
storing each row of the matrix, its only stores rows that have
non-zero elements. 
To this end, uCSR uses an \emph{unordered} array \code{m_idx} to store
the positions of non-empty rows. 
Those rows are in turn compressed using SV, with associated
two-dimensional arrays \code{m_ind} and \code{m_val} storing indices
and values of non-empty cells for each row.\footnote{In a more
  realistic encoding, an additional dense array \fcode{m_ptr} is used
  to store locations of the boundaries between rows in
  \emph{one-dimensional} arrays \fcode{m_val} and \fcode{m_ind}. In
  this section, for simplicity of the presentation, we treat
  \fcode{m_ind} and \fcode{m_val} as \emph{two-dimensional} arrays. In
  our mechanisation, however, we verify the encoding involving the
  \fcode{m_ptr} array (\cf~\autoref{sec:eval}).}
This intuition is mirrored in the programmatic specification of uCSR
shown in~\autoref{fig:ucsr}, which takes three arrays \code{m_idx},
\code{m_ind}, and \code{m_val}, as well as the dense row/column
indices \code{i} and \code{j}, and retrieves the corresponding element
by calling \code{sv_get} on its actual row encoding.



\begin{wrapfigure}[19]{r}{0.51\textwidth}
\vspace{-13pt}
  \begin{minted}[numbers=left, xleftmargin=2.5ex, numbersep=6pt, escapeinside=@@, fontsize=\footnotesize]{python3}
  def spmspv(m_idx, m_ind, m_val, x_ind, x_val):
    ans = malloc(M)
    for i in range(0, length(m_idx)):
      ans[m_idx[i]] = 
        spvspv(m_ind[i], m_val[i], x_ind, x_val)
    return ans

  def spvspv(y_ind, y_val, x_ind, x_val):
    s, iY, iX = 0, 0, 0
    while (iY < length(y_ind)) && 
          (iX < length(x_ind)):
      if y_ind[iY] = x_ind[iX]:
        s += y_val[iY] * x_val[iX]
        iY++; iX++
      else if y_ind[iY] < x_ind[iX]:
        iY++
      else if x_ind[iX] < y_ind[iY]:
        iX++
    return s
  \end{minted}
  \setlength{\abovecaptionskip}{7pt}  
  \caption{Sparse Matrix/Vector product, SpMSpV (top) and Sparse
    Vector/Vector product, SpVSpV (bottom).}
  \label{fig:smpv}
\end{wrapfigure} 
The implementation of the sparse matrix/vector product is given
in~\autoref{fig:smpv}; we have adopted it from the output produced by
the \tname{Finch} compiler~\cite{AhrensDKA23}.
The function \code{spmspv} takes a sparse matrix~$m$, encoded in uCSR
format via three arrays: \code{m_idx}, \code{m_ind}, and \code{m_val},
as well as a sparse vector~$x$, encoded via \code{x_ind} and
\code{x_val}. 
Its result is a dense array \code{ans}, populated element-wise by
the components of the dot-product $m \cdot x$.
The implementation of \code{spmspv}, also uses the height \code{M} of
the matrix, which we assume to be a global variable.
The body of \code{spmspv} first iterates over all non-zero rows in the
order determined by \code{m_idx} (line 3). At each iteration of the
\code{for}-loop it calls \code{spvspv}, computing the product of two
sparse vectors, \code{m_idx[i]}\ith row of the matrix~$m$ and the
vector~$x$.
The \code{while}-loop in \code{spvspv} (lines~10-18)
\emph{co-iterates} over a pair of sparse vectors, visiting only
non-zero values of \emph{both} arrays.
It does so by maintaining the positions in the corresponding vectors
via variables \code{iY} and \code{iX}, advancing them in a way so that
the result stored in~\code{s} is changed only when indices of non-zero
elements in the two vectors coincide (line~12). 



To specify the result of \code{spmspv}, let us introduce some
notation.
We will abbreviate all unmodified arrays that encode the
involved data by the following Separation Logic (SL)-style predicates:
\begin{equation}
\label{eq:arrays}
{\small{
\begin{array}{rcl}
  \arrs_{\texttt{m}} &\eqdef& \arr{\tinds{m}}{\inds{m}}\str \arr{\tvals{m}}{\vals{m}} \str \arr{\tidxs{m}}{\idxs{m}}\\
  \arrs_{\texttt{x}} &\eqdef& \arr{\tinds{x}}{\inds{x}} \str \arr{\tvals{x}}{\vals{x}} \\ 
  \arrs &\eqdef& \arrs_{\texttt{m}} \str \arrs_{\texttt{x}}
\end{array}
}}  
\end{equation}
That is, for instance, $\arrs_{\texttt{m}}$ combines three arrays
responsible for representing a sparse matrix $m$, with their base
pointers (\eg, $\tinds{m}$) given in \texttt{monospace} font and
payloads (\eg, $\inds{m}$) given in \emph{italic}.
The disjointness of the arrays is asserted by the SL \emph{separating
  conjunction} connective ($\str$).

The desired specification for \code{spmspv} is, therefore, captured by
the following hypersafety triple:
\vspace{-3pt}
\begin{equation}
\label{eq:spmspv1}
{\small{
\!\!\!
\hspec{\!\!\!
\begin{array}{c}
  \\
  \arrs(\cdot)
  \\
  \phantom{k}
\end{array}
\!\!\!}    
\hmid{\!\!\!
  \begin{array}{c@{\ }c@{\ }l}
  1 &:& \mathtt{spmspv(m_{idx}, m_{ind}, m_{val}, x_{ind}, x_{val})} 
  \\ [2pt]
  \angled{2, i, j}_{\!\!\!\!\tiny
    \begin{array}{l} i \in [0, M) \\ j \in [0,N) \end{array}} &:&  \mathtt{ucsr\_get(m_{idx}, m_{ind}, m_{val},} i, j)
  \\ [2pt]
  \angled{3, j}_{j \in [0, N)} &:& \mathtt{sv\_get(x_{ind}, x_{val},} j)
  \end{array}
\! \!\!}
\hspec{\!\!\!
\left.
\begin{array}{r}
a \\[2pt] \many{m} \\[2pt] \many{x}
\end{array}
\right|\!\!
\begin{array}{l}
\exists s, \arrl{a\code{(1)}}{s} \str
\\[2pt]
\pure{\forall i, s[i] = \sum\limits_j \many{m}(i,j)\many{x}(j)}~\str  
\\[2pt]
\arrs(\cdot)
\end{array}
\!\!\!}
}}
\end{equation}
The spec assumes that the width $N$ of the matrix $m$ is a global
variable.
It makes use of the ($\cdot$) notation that ``replicates'' a SL across
all state components in the pre/postcondition.
The $\pure{\cdot}$ notation stands for \emph{pure} assertions that
constrain values and do not depend on the shape of the heaps.
Notice that, unlike $\arrs(\cdot)$, the \emph{spatial} part
$\small{\arrl{a\code{(1)}}{s}}$ is only present in the first component's
post-state.

\paragraph{What's next?}
In the rest of this section, we will verify the validity of the
specification~\eqref{eq:spmspv1} via the rules of \logic.
We will start by focusing on the \code{for}-loop from the
\code{spmspv}'s body, ``peeling away'' all calls to \code{ucsr_get},
that do not correspond to any of the loop's iterations
(\autoref{sec:seq-prod}~--~\autoref{sec:nest-frame}).
We will then split the remaining calls into groups (one group per
non-zero row) and relate the result of each one to one iteration of
the \code{for}-loop (\autoref{sec:for}).
We will then show how the obtained goal can be proven
\emph{compositionally} by using an independently verified
specification of \code{spvspv} (\autoref{sec:subst-spec}).
Finally, we will sketch the proof of \code{spvspv}, showcasing
\logic's rule for \code{while}-loops (\autoref{sec:while}).

\subsection{Proofs about Sequential Composition and Independent
  Programs}
\label{sec:seq-prod}

\begin{figure}[t]
    \begin{minipage}{0.48\textwidth}
      {\small{
  \begin{mathpar}
  \!\!\!\!\!\!\!\!\!\!\!\!
  \inferrule[\tname{SeqU1}]
  {
  \begin{array}{c}
    \hsll{\! P\! }{\iota:\ p_1}{{x}}{H \!\! }
    \arcr[2pt]
    \hsllazy{\! H \ \ }{\iota:\ p'_1, \ S:\ \pp_2\ }{{x}, \many{z}}{Q(\many{z}) \!\! }    
  \end{array}
  }
  {
  \hsllazy{\! P\ \ }{\iota : p_1; p'_1,~~S : \pp_2\ }{{x},
    \many{y}}{Q(x\many{y}) \!\! }
  }
  \end{mathpar}
  }}
  \end{minipage}
  \!\!\!\!
  \begin{minipage}{0.48\textwidth}
    {\small{  
  \inferrule[\tname{SeqU2}]
  {
  \begin{array}{c}
  \hsllazy{\!\! P \ }{\iota:\ p_1,\ S:\ \pp_2\ }{{x}, \many{z}}{H(\many{z}) \!\! }
  \arcr[2pt]
  \forall \many{z}, \hsll{\! H(\many{z})\! }{\iota:\ p'_1}{{x}}{Q({x}\many{z}) \!\! }
  \end{array}
  }
  {
  \hsllazy{\! P\ \ }{\iota : p_1; p'_1,~~S : \ \pp_2\ }{{x_1}, \many{x_2}}{Q({x_1}\many{x_2}) \!\! }
  }  
  }}
  \end{minipage}
  \vspace{2pt}
  \begin{minipage}{\textwidth}
      {\small{
        \begin{mathpar}
          \inferrule*[Left=\tname{Product}]
          {
            \local{\{P_i, Q_i\}}{i} \qquad
            \forall i, \hsll{\pure{i \in S} \str P_i}{i:\ \pp(i)}{x}{Q_i(x)}     
          }
          {
          \hsll{\bigstr_{i\in S} P_i}{S:\ \pp}{\many{x}}{\bigstr_{i\in S} Q_i(\many{x}(i))}
          }
          \end{mathpar} 
    }}
    \end{minipage}
    \caption{Rules for sequential composition and independent
      programs.}
\label{fig:seq}
\end{figure}

To begin with the proof of~\eqref{eq:spmspv1}, we unfold the definition
of \code{spmspv} in the first component, and omit ranges of variables
$i$ and $j$, obtaining the following goal to prove:
\begin{equation}
  \label{eq:spmspv2}
  {\small{
  \hspec{\!\!\!
  \begin{array}{c}
    \\
    \ \arrs(\cdot) \ \ \ 
    \\
    \phantom{k}
  \end{array}
  \!\!\!}    
  \hmid{\!\!\!
    \begin{array}{c@{\ }c@{\ }l}
    1 &:& \Ans = \kw{malloc}\code{(M);}\dots
    \\ [2pt]
    \angled{2, i, j} &:&  \mathtt{ucsr\_get(m_{idx}, m_{ind}, m_{val},} i, j)
    \\ [2pt]
    \angled{3, j} &:& \mathtt{sv\_get(x_{ind}, x_{val},} j)
    \end{array}
  \!}
  \hspec{\!\!\!
  \left.
  \begin{array}{r}
  a \\[2pt] \many{m} \\[2pt] \many{x}
  \end{array}
  \right|
  \begin{array}{l}
  \exists s, \arrl{a\code{(1)}}{s} \str
  \\[2pt]
  \pure{\forall i, s[i] = \sum_j \many{m}(i,j)\many{x}(j)} 
  \\[2pt]
  \str \arrs(\cdot) 
  \end{array}
  \!\!\!}
  }}
\end{equation}
Notice that the \code{for}-loop at lines~3-5 of \autoref{fig:smpv}
only ``visits'' rows of $m$ that correspond to \code{m_idx[i]} for
some \code{i}.
In this subsection, we will show how to reduce the verification
goal~\eqref{eq:spmspv2} to the one that only mentions those
\code{ucsr_get}-components whose (row) argument $i$ is present as an
element in \code{m_idx}.


\subsubsection{Rules for Sequential Composition}
\label{sec:seq}

Our first step is to get to the \code{for}-loop in the first component
of the goal~\eqref{eq:spmspv2}, which will require us to symbolically
step through the \code{malloc} statement, while ``focusing'' on the
corresponding component.
This focusing capability is achieved by the \rulename{SeqU1} shown in
\autoref{fig:seq}, which allows one to split the index set of a
hyper-triple into two disjoint subsets, $\set{\iota}$ and $S$,
symbolically executing the first component $p_1$ of the sequential
composition $p_1;p'_1$ that corresponds to the $\iota$-indexed
component of the entire hyper-triple.\footnote{To avoid clutter, the
  rules assume that execution of the ``projected'' components, such as
  $\iota : p_1$ in \rulename{SeqU1}, are implicitly \emph{framed} with
  the corresponding hyper-heap footprint of the $S$-indexed component,
  which it keeps unmodified.}
Applying \rulename{SeqU1} with $\iota = 1$ to the
goal~\eqref{eq:spmspv2} allows us to execute the \code{malloc}
statement using the proof rule for allocation, which is relatively
standard for Separation Logic-style formalisms, so we postpone its
presentation until \autoref{sec:logic}.
%
%
%
The first premise of \rulename{SeqU1}, thus, takes the form of the
following goal, where we use $\Zero$ to denote a 0-filled vector of a suitable
size, evident from the context:
\begin{equation*}
\label{eq:malloc}
  {\small{
\hspec{
  \arrs(\cdot)
}
\hmid{
\begin{array}{c}
 1 : \Ans = \kw{malloc}\code{(M)}
\end{array}
}
\hspec{
\arrs(\cdot)~\str~\arr{ans\code{(1)}}{\mathbf{0}}
}
}}
\end{equation*}
Here the $\code{(1)}$ syntax in $\small{\arr{ans\code{(1)}}{\textbf{0}}}$
means that \code{ans} is a pointer allocated in the 1-indexed
component of the symbolic state (\ie, the heap where the first
component runs).
Abbreviating the \code{for}-loop as~$f$ we obtain the following triple
as the result of the second obligation of \rulename{SeqU1}:
\begin{equation}
  \label{eq:spmspv22}
  {\small{
  \hspec{\!\!\!
  \begin{array}{c}
    \arr{ans(1)}{\mathbf{0}}
    \\
    \str \arrs(\cdot)
  \end{array}
  \!\!\!}    
  \hmid{\!\!\!
    \begin{array}{c@{\ }c@{\ }l}
    1 &:& f;\ \kw{return}\code{ ans}
    \\ [2pt]
    \angled{2, i, j} &:&  \mathtt{ucsr\_get(m_{idx}, m_{ind}, m_{val},} i, j)
    \\ [2pt]
    \angled{3, j} &:& \mathtt{sv\_get(x_{ind}, x_{val},} j)
    \end{array}
  \!}
  \hspec{\!\!\!
  \left.
  \begin{array}{r}
  a \\[2pt] \many{m} \\[2pt] \many{x}
  \end{array}
  \right|
  \begin{array}{l}
  \exists s, \arrl{a\code{(1)}}{s} \str
  \\[2pt]
  \pure{\forall i, s[i] = \sum_j \many{m}(i,j)\many{x}(j)} 
  \\[2pt]
  \str \arrs(\cdot) 
  \end{array}
  \!\!\!}
  }}
\end{equation}
The goal is further simplified by stripping off the \code{return}
statement via the \rulename{SeqU2} and unfolding
$\arr{ans(1)}{\textbf{0}}$ into an iterated separating conjunction
$\bigstr_{i=0}^M \mcode{ans(1)}+i\mapsto 0$ over a set of
\emph{points-to heaplets} (\ie, symbolic heap entries), indexed by
their offsets $i$ from the base array address $\mcode{ans(1)}$:
\begin{equation*}
  {\small{
  \hspec{\!\!\!
  \begin{array}{c}
    \bigstr_{i=0}^M \code{ans(1)}+i\mapsto 0
    \\
    \str \arrs(\cdot)
  \end{array}
  \!\!\!}    
  \hmid{\!\!\!
    \begin{array}{c@{\ }c@{\ }l}
    1 &:& \forl{i}{0}{|\idxs{m}|}{p(i)}
    \\ [2pt]
    \angled{2, i, j} &:&  \mathtt{ucsr\_get(m_{idx}, m_{ind}, m_{val},} i, j)
    \\ [2pt]
    \angled{3, j} &:& \mathtt{sv\_get(x_{ind}, x_{val},} j)
    \end{array}
  \!}
  \hspec{\!\!\!
  \left.
  \begin{array}{r}
  - \\[2pt] \many{m} \\[2pt] \many{x}
  \end{array}
  \right|
  \begin{array}{l}
  \exists s, \bigstr_{i=0}^M \code{ans(1)}+i\mapsto s[i] \str
  \\[2pt]
  \pure{\forall i, s[i] = \sum_j \many{m}(i,j)\many{x}(j)} 
  \\[2pt]
  \str \arrs(\cdot) 
  \end{array}
  \!\!\!}
  }}
\end{equation*}
Finally, we abbreviate the loop's body as $p(i)$, and simplify the
postcondition by substituting each value of the \code{ans}' payload
sequence $s[i]$ with the correspondent sum expression:
\begin{equation}
  \label{eq:spmspv33}
  {\small{
  \hspec{\!\!\!
  \begin{array}{c}
    \bigstr_i \code{ans(1)}+i\mapsto 0
    \\
    \str \arrs(\cdot)
  \end{array}
  \!\!\!}    
  \hmid{\!\!\!
    \begin{array}{c@{\ }c@{\ }l}
    1 &:& \forl{i}{0}{|\idxs{m}|}{p(i)}
    \\ [2pt]
    \angled{2, i, j} &:&  \mathtt{ucsr\_get(m_{idx}, m_{ind}, m_{val},} i, j)
    \\ [2pt]
    \angled{3, j} &:& \mathtt{sv\_get(x_{ind}, x_{val},} j)
    \end{array}
  \!}
  \hspec{\!\!\!
  \left.
  \begin{array}{r}
  - \\[2pt] \many{m} \\[2pt] \many{x}
  \end{array}
  \right|
  \begin{array}{l}
  \bigstr_i\code{ans(1)}+i\mapsto \\
  \qquad\qquad \sum_j \many{m}(i,j)\many{x}(j)
  \\[2pt]
  \str \arrs(\cdot) 
  \end{array}
  \!\!}
  }}
\end{equation}

\subsubsection{Reasoning with Component-Local Assertions}
\label{sec:prod} 

%
Remember that array $\tidxs{m}$ contains indices of non-zero rows of
the matrix $m$, so that the \code{for}-loop from the first component
will iterate only over such indices. 
This suggests us to first get rid of all the components with
$\mathtt{ucsr\_get(m_{idx},m_{ind},m_{val}},i,j)$ for
$i \notin \idxs{m}$, and try to relate the rest of those with each
iteration of the \code{for}-loop. 
To implement the former, we note that for each index $i$ outside of
the payload $\idxs{m}$ the result of
$\mathtt{ucsr\_get(m_{idx},m_{ind},m_{val}},i,j)$
will be~0 for any $j$.
This fact can be stated as the following triple:
\begin{equation}
  \label{eq:spmspv4}
  {\small{
  \hspec{\!\!\!
  \begin{array}{c}
    \arrs(\cdot)
  \end{array}
  \!\!\!}    
  \hmid{\!\!\!
    \begin{array}{c@{\ }c@{\ }l}
    \angled{2, i, j}_{\!\!\!\!\tiny
    \begin{array}{l} i \not\in \idxs{m} \\ j \in [0,N) \end{array}} &:&  \mathtt{ucsr\_get(m_{idx}, m_{ind}, m_{val},} i, j)
    \end{array}
  \!}
  \hspec{\!\!\!
  \begin{array}{r}
  ~~\many{m}
  \end{array}
  \!\!\left|
  \begin{array}{l}
  \pure{\forall i \notin \idxs{m}\, \forall j, \many{m}(i,j) = 0}
  \\[2pt]
  \str \arrs(\cdot) 
  \end{array}
  \right.
  \!\!\!}
  }}
\end{equation}
To see how to deal with such triples, let us consider the following
simple verification goal:
\begin{equation*}
{\small{
  \hsll{s(\cdot)\mapsto y}{i_{0 \leq i < N}:\mathtt{s}\
    {\text{\fontsize{9}{18}\selectfont +=}}\ i }{\_}{s(\cdot)\mapsto y
    + i}
}}
\end{equation*}
Intuitively, proving this triple is equivalent to proving~$N$
independent 1-safety triples of the form
\begin{equation*}
{\small{
  \hsll{s(i)\mapsto y}{i:\mathtt{s}\
    {\text{\fontsize{9}{18}\selectfont +=}}\ i }{\_}{s(i)\mapsto y +
    i}
}}
\end{equation*}
To capture this intuition \logic features the rule \rulename{Product}
(\cf \autoref{fig:seq}) that allows one to divide the state assertions
in the pre-/postcondition (indexed by a set $S$) into $|S|$ disjoint
assertions, each being \emph{local} to a particular index.
The idea of locality comes from the \logic memory model, so that its
state assertions constrain not just the shapes of the heap but also
the component in which they reside, so that, \eg, $x(1) \mapsto 0$ and
$x(2) \mapsto 0$ are not satisfied by the same heap.
Postponing a more formal treatment of \logic state till
\autoref{sec:logic}, here we say that an \logic assertion is local to
an index~$\iota$ if it can only be satisfied by states that belong to
the $\iota\th$ component.
With the \rulename{Product} rule, the assumed locality of the
individual specifications and independence of the program runs, it
suffices to prove the correspondent triple for each individual
program~$p$ from a hyper-triple.

The postcondition of the specification~\eqref{eq:spmspv4} can be
rewritten using the following equivalence:
\begin{equation*} {\small{ \pure{\forall i \notin \idxs{m}\, \forall
        j, \many{m}(i,j) = 0} \Longleftrightarrow \bigstr_{i \notin \idxs{m}, j}
      \pure{\many{m}(i,j) = 0} } }
\end{equation*}
With the right-hand side of the equivalence above, we can apply the
\tname{Product} rule, reducing~\eqref{eq:spmspv4} to
\begin{equation}
  \label{eq:spmspv6}
  {\small{
  \hspec{\!\!\!
  \begin{array}{c}
    \pure{i \not\in \idxs{m} \land j \in [0,N)} \str\\[2pt]
    \arrs(\cdot)
  \end{array}
  \!\!\!}    
  \hmid{
  \angled{2,i,j}:  \mathtt{ucsr\_get(m_{idx}, m_{ind}, m_{val},} i, j)
  }
  \hspec{\!\!\!
  \begin{array}{r}
  m
  \end{array}
  \left|
  \begin{array}{l}
    \pure{m=0} \str \\[2pt]
    \arrs(\cdot)  
  \end{array}
  \right.
  \!\!\!}
  }}
\end{equation}
%
%
This is a 1-safety property, which is stated in terms of a traditional
Hoare-style triple of the form $\hsll{P}{\iota : p}{x}{Q(x)}$for a
program~$p$.
To discharge such obligations, \logic provides the standard set of
Separation Logic rules. We omit this proof, which relies on an
independently proven specification of $\mathtt{index(m_{idx},}i)$; the
reader can find it in our Coq development~\cite{lgtm}.

\subsection{Focusing and Framing}
\label{sec:nest-frame}
 
At this point, one can think of \logic as of a ``Separation Logic
going hyper'', with the heaps being replicated over index sets.
An astute reader could have noticed that the rule \rulename{Product}
exploits the (hyper-)locality of spatial triples to provide a version
of SL's \emph{frame} rule that combines multiple independent
\emph{unary triples} into a single \emph{hyper-triple}.
In our next steps, we will keep taking advantage of the SL nature of
\logic that enables principles of modular reasoning about
hypersafety.

\begin{wrapfigure}[7]{r}{0.42\textwidth}
  \setlength{\abovecaptionskip}{5pt}
  \vspace{-18pt}
  \begin{minipage}{0.9\linewidth}
  {\small{  
  \centering
  \begin{mathpar}
    \inferrule[\tname{Focus}]
    {
      \begin{array}{c}
        S = S_1 \uplus S_2 \arcr[2pt]
        \hsll{\! P\! }{S_1:\ \pp_1}{\many{x}}{H(\many{x}) \!\! } \arcr[2pt]
        \forall \many{x},\ \hsll{\! H(\many{x}) \! }{S_2:\ \pp_2}{\many{y}}{Q(\many{x}\many{y}) \!\! }
      \end{array}
    }
    {
      \hsll{\! P\! }{S:\ \pp_1 \uplus \pp_2}{\many{z}}{Q(\many{z}) \!\ }
    }
  \end{mathpar}
 }}
  \end{minipage}
\caption{Focus rule}
\label{fig:nest}
\end{wrapfigure}
Using~\eqref{eq:spmspv4}, we can remove all runs of \code{ucsr_get}
for $m$'s zero-only rows of $m$ from the goal~\eqref{eq:spmspv33}.
To do so, we will make use the \rulename{Focus} rule, shown in
\autoref{fig:nest}. Applying it, followed by the framing-out of some
no longer useful pre-/postconditions components (via standard SL
framing rules), will make us ready to handle the \code{for}-loop in
the first component.
The \tname{Focus} rules allows us to advance executions of the programs
within $S_1$ (a subset of the whole index set $S = S_1 \uplus S_2$),
keeping the rest of the state unchanged. 
The notation $\pp_1 \uplus \pp_2$ stands for the disjoint union of two
sets of indexed programs.
Notice that the range of the result vector $\many{y}$ of $S_2: \pp$
would be $S_2$, so we will extend it to the whole $S$ by appending
with the result of $S_1: \pp$, vector $\many{x}$ of range $S_1$. 
We
apply the \tname{Focus} rule to the goal~\eqref{eq:spmspv33} by taking
$S_1 = \set{\angled{2, i, j} \mid {i \notin \idxs{m}}}$ and $S_2 =
\set{1} \cup \set{\angled{2, i, j} \mid {i \in \idxs{m}}} \cup
\set{\angled{3, j} \mid \ldots}$.
Using the spec~\eqref{eq:spmspv4} on the $S_1$-indexed part earns us
the conjunct
$ \pure{\forall i \notin \idxs{m}\, \forall j, \many{m}(i,j) = 0}$ in
the precondition of the following goal, which corresponds to the
second premise of the \rulename{Focus}:
\begin{equation*}
  \label{eq:spmspv3}
  {\small{
  \hspec{\!\!\!
  \begin{array}{c}
    \bigstr\limits_i \code{ans(1)}+i\mapsto 0 \\[-2pt]
    \str \arrs(\cdot)
  \end{array}
  \!\!\!}    
  \hmid{\!\!\!
    \begin{array}{c@{\ }c@{\ }l}
    1 &:& \forl{i}{0}{|\idxs{m}|}{p(i)}
    \\ [2pt]
    \angled{2, i, j}_{i \in \idxs{m}} &:&  \mathtt{ucsr\_get(m_{idx}, m_{ind}, m_{val},} i, j)
    \\ [2pt]
    \angled{3, j} &:& \mathtt{sv\_get(x_{ind}, x_{val},} j)
    \end{array}
  \!}
  \hspec{\!\!\!
  \begin{array}{r}
  - \\[2pt] \many{m} \\[2pt] \many{x}
  \end{array}
  \left|
  \begin{array}{l}
  \bigstr\limits_{i \in \idxs{m}}\!\!\code{ans(1)}+i\mapsto \\[-6pt]
  \qquad\quad \sum_j \many{m}(i,j)\many{x}(j)
  \\[2pt]
  \bigstr\limits_{i \notin \idxs{m}}\!\!\code{ans(1)}+i\mapsto 0
  \\[2pt]
  \str \arrs(\cdot) 
  \end{array}
  \right.
  \!\!}
  }}
\end{equation*}
Now, we can once again exploit the Separation Logic capabilities of
\logic, framing out the part
{\small{
\(\bigstr_{i\not\in \idxs{m}} \code{ans(1)} + i \mapsto 0\)}}
from both pre- and postconditions, obtaining the following goal:
\begin{equation}
  \label{eq:spmspv8}
  {\small{
\!\!\!
  \hspec{\!\!\!
  \begin{array}{c}
    \bigstr\limits_{i \in \idxs{m}}\!\!\!\code{ans(1)}+i\mapsto 0 \\[-2pt]
    \str \arrs(\cdot)
  \end{array}
  \!\!\!}    
  \hmid{\!\!\!
    \begin{array}{c@{\ }c@{\ }l}
    1 &:& \forl{i}{0}{|\idxs{m}|}{p(i)}
    \\ [2pt]
    \angled{2, i, j}_{i\in\idxs{m}} & :&  \mathtt{ucsr\_get(m_{idx}, m_{ind}, m_{val},} i, j)
    \\ [2pt]
    \angled{3, j} &:& \mathtt{sv\_get(x_{ind}, x_{val},} j)
    \end{array}
  \!\!\!}
  \hspec{\!\!\!
  \left.
  \begin{array}{r}
  - \\[2pt] \many{m} \\[2pt] \many{x}
  \end{array}
  \right|\!\!
  \begin{array}{l}
  \bigstr\limits_{i \in \idxs{m}}\!\!\!\code{ans(1)}+i\mapsto \\[-6pt]
  \qquad\quad \sum_j \many{m}(i,j)\many{x}(j)
  \\[2pt]
  \str \arrs(\cdot) 
  \end{array}
  \! \!\!\!}
  }}
\end{equation}

\noindent
The triple~\eqref{eq:spmspv8} is an examplary case of a parametrised
hypersafety specification: its arity is determined by the data
$\idxs{m}$ that serves as a selector for the \code{ucsr_get} programs.
Phrasing it this way allowed us to bring the iterations of
\code{spmspv}'s \code{for}-loop in line with the indexing structure of
the specification components, so we could handle the remaining proof
using a dedicated \logic rule for loops.

\subsection{Iteration and Array Splitting}
\label{sec:for}


To prove the goal~\eqref{eq:spmspv8}, we observe that the
$i$\ith~\code iteration of the {for}-loop in the first component
computes the dot product of the vector $x$ and the $i$\ith row of the
matrix $m$.
This suggest partitioining the verification task by \emph{aligning}
(\ie, relating the result of) each $i$\ith iteration of that loop with
a sequence of runs of $\mathtt{sv\_get(x_{ind}, x_{val},} j)$ and
$\mathtt{ucsr\_get(m_{idx}, m_{ind}, m_{val},} i, j)$ for all $j$,
$0 \leq j < N$, as those runs compute values of $x$ and the
$i$\ith row of $m$.
This pattern of splitting a set of independent runs into ``batches''
to align with individual loop iterations in the ``main'' program is so
common in structured data computations that \logic introduces a
dedicated rule \rulename{For} to handle it (\autoref{fig:forsimpl}).

To understand how the rule works, first notice its premise
$S = \bigcup_{i=0}^{N - 1} S_i$ that splits the index set $S$ into $N$
parts $S_i$. Each part corresponding to a program ``batch'' to be
aligned with one iteration of the \code{for}-loop.
The precondition of the conclusion features an assertion
$ \bigstr_{i=0}^{N - 1} R_i$ that describes a matching split of the
pre-state required to run each of those batches (hence the locality
side condition).
The most interesting aspect of the rule is the \emph{loop
  invariant}~$I$, which is an assertion parametrised by ($i$)~an
integer that corresponds to the \emph{next} value of the loop counter
and ($ii$)~a hyper-value that collects the results of all program
batches that have been already aligned with the ``prefix'' of the
loop.
That is, in the conclusion of the rule, it is assumed that the
assertion $I(0, \many{\emptyset})$ holds initially: which corresponds
to zero iterations and no aligned batches executed so far
($\many{\emptyset}$ is an empty hyper-value).
At the end, $I(N, \many{y})$ means that the loop has been fully
executed and all of the aligned programs' results are bound
by~$\many{y}$.
The inductive step of the rule, \ie, the premise defining the goal for
the body of the loop~$p$, assumes that the invariant initially holds
for some $i < N$ and a hyper-value $\many{x}$ (of size $i$, which we
omit from the rule for brevity), and asserts in the postcondition that
at the end the invariant holds on the next index $i + 1$ and the
combined hyper-value $\many{x}\many{y}$.%
\footnote{As a curiosity, this means that $I$ has a dependent type:
  the value of its first parameter determines the arity of the
  second.}

\begin{figure}
    \begin{minipage}{1.0\linewidth}
    {\small{
    \begin{mathpar}
    \!\!\!\!
    \inferrule*[right=\tname{For}]
    {
    \begin{array}{c}
    S = \bigcup_{i=0}^{N - 1} S_i
    \quad
    \local{R_i}{S_i}
    \quad
    \forall \many{x}, i < N,
          \hsllazy{
            I(i, \many{x}) \str
            R_i \ \ }
            {\iota : p,~~S_i : \pp_i}
            {z, \many{y}}
            {I(i+1, \many{x}\many{y})}
    \end{array}
    }
    {
    \hsllazy{
            I(0, \many{\emptyset}) \str
            \bigstr_{i=0}^{N - 1} R_i \ \ }
            {\iota : \forl{i}{0}{N}{p},\
             S : \pp}
            {z, \many{y}}
            {I(N, \many{y}) }}
    \end{mathpar}
    }}
    \end{minipage}
    \caption{A rule for \kw{for}-iteration.}
    \label{fig:forsimpl}
\end{figure}

To proceed with our proof of~\eqref{eq:spmspv8} using the
\rulename{For} rule, we first split the set of its indices as follows:
\begin{equation}
  \label{eq:ri}
\!\!\!
{\small{\angled{2,i,j}_{i \in \idxs{m}, j \in [0,N)} \cup \angled{3,
        j}_{j \in [0, N)}= \bigcup_{i \in 0}^{|\idxs{m}|-1} S_i = \bigcup_{i \in 0}^{|\idxs{m}|-1}
      \left(\angled{2,\idxs{m}[i],j}_{j \in [0,N)} \cup \angled{3, j}_{j \in
        [0, N)}\right)
}}
\end{equation}
Next, we define the loop invariant as follows:
\begin{equation}
  \label{eq:inv}
  {\small{
  \begin{array}{rcl}
  \Ispmspv(i, \many{mx}) & \eqdef &
  \begin{array}{l}
  \bigstr_{j=0}^{i-1} \code{ans(1)} + j \mapsto \sum_{k=0}^N \many{m}(j,k)\many{x}(k) \str
  \bigstr_{j = i}^{|\idxs{m}| - 1} \code{ans(1)} + j \mapsto 0
  \end{array}
  \end{array}
  }}
\end{equation}

\noindent
The invariant states that (a) a $j$\ith element from the prefix of the
array $\mathtt{ans}(1)$, is already filled with the dot product of $x$
and $j$\ith row of $m$ (\ie, the dot-product of results of
\code{ucsr_get} and \code{sv_get}), for $0 \leq j < i$ and (b) that
the suffix of the array is filled with zeros.
After applying \rulename{For} and advancing the code of \code{spmspv}
(\autoref{fig:smpv}) up to the line~5, we obtain the following
goal, for some $i \in [0, |\idxs{m}|)$:
\begin{equation*}
  \label{eq:spvspv_uscr_get}
  {\small{
  \hspec{\!\!\!\!\!\!\!\!\!\!\!\!
  \begin{array}{c}
    \code{ans(1)}~+ \\ \qquad \idxs{m}[i]\mapsto 0 \\[2pt]
    \str \arrs(\cdot)
  \end{array}
  \!\!\!}    
  \hmid{\!\!\!
    \begin{array}{c@{\ }c@{\ }l}
    1 &:& \mathtt{ans[m_{idx}[i]]=spvspv(\dots)}
    \\ [2pt]
    \angled{2, i, j}_{j\in [0,N)} &:&  \mathtt{ucsr\_get(m_{idx}, m_{ind}, m_{val},} \idxs{m}[i], j)
    \\ [2pt]
    \angled{3, j}_{j\in [0,N)} &:& \mathtt{sv\_get(x_{ind}, x_{val},} j)
    \end{array}
  \!}
  \hspec{\!\!\!
  \left.
  \begin{array}{r}
  - \\[2pt] \many{m} \\[2pt] \many{x}
  \end{array}
  \right|
  \begin{array}{l}
  \code{ans(1)}+\idxs{m}[i]\mapsto \\
  \qquad\quad \sum_j \many{m}(i,j)\many{x}(j)
  \\[2pt]
  \str \arrs(\cdot) 
  \end{array}
  \!\!}
  }}
\end{equation*}
Now we can advance lines~2 and 3 in
\code{uscr_get}~(\autoref{fig:ucsr}):
$\mathtt{\mcode{index}(m_{idx}},\idxs{m}[i])$ evaluates to $i$,
triggering then-branch of the \code{if}-statement (line~4), leaving us
with the following goal:
\begin{equation}
  \label{eq:spvspv_dom1}
  {\small{
  \hspec{\!\!\!\!\!\!\!\!\!\!\!\!
  \begin{array}{c}
    \code{ans(1)}+ \\ \qquad \idxs{m}[i]\mapsto 0 \\[2pt]
    \str \arrs(\cdot)
  \end{array}
  \!\!\!}    
  \hmid{\!\!\!
    \begin{array}{c@{\ }c@{\ }l}
    1 &:& \mathtt{ans[m_{idx}[i]]=spvspv(\dots)}
    \\ [2pt]
    \angled{2, i, j}_{j\in [0,N)} &:&  \mathtt{sv\_get(m_{ind}[i], m_{val}[i],} j)
    \\ [2pt]
    \angled{3, j}_{j\in [0,N)} &:& \mathtt{sv\_get(x_{ind}, x_{val},} j)
    \end{array}
  \!}
  \hspec{\!\!\!
  \left.
  \begin{array}{r}
  - \\[2pt] \many{m} \\[2pt] \many{x}
  \end{array}
  \right|
  \begin{array}{l}
  \code{ans(1)}+\idxs{m}[i]\mapsto \\
  \qquad\quad \sum_j \many{m}(i,j)\many{x}(j)
  \\[2pt]
  \str \arrs(\cdot) 
  \end{array}
  \!\!}
  }}
\end{equation}
We can further reduce the goal using \rulename{SeqU1} and omitted
symbolic execution rules for unary triples:
\begin{equation}
  \label{eq:spvspv_dom}
  {\small{
  \hspec{\!\!\!
  \begin{array}{c}
    \phantom{k} \\
    \arrs(\cdot) \\ 
    \phantom{k}
  \end{array}
  \!\!}    
  \hmid{\!\!\!
    \begin{array}{c@{\ }c@{\ }l}
    1 &:& \mathtt{spvspv(m_{ind}[i], m_{val}[i], x_{ind}, x_{val})}
    \\ [2pt]
    \angled{2, i, j}_{j\in [0,N)} &:&  \mathtt{sv\_get(m_{ind}[i], m_{val}[i],} j)
    \\ [2pt]
    \angled{3, j}_{j\in [0,N)} &:& \mathtt{sv\_get(x_{ind}, x_{val},} j)
    \end{array}
  \!}
  \hspec{\!\!\!
  \left.
  \begin{array}{r}
  s \\[2pt] \many{m} \\[2pt] \many{x}
  \end{array}
  \right|
  \begin{array}{l}
  \pure{s= \sum_j \many{m}(i,j)\many{x}(j)}
  \\[2pt]
  \str \arrs(\cdot) 
  \end{array}
  \!\!}
  }}
\end{equation}

\subsection{Domain Substitution and Specification Composition}
\label{sec:subst-spec}

At this point, we have \emph{almost} reduced the verification of the
\mcode{spmspv} specification to verifying the correctness of its
auxiliary function \mcode{spvspv} \wrt the respective format
definition \mcode{sv_get}: our remaining goal~\eqref{eq:spvspv_dom}
looks pretty much like a specification of a sparse vector/vector
dot-product.

The actual desired specification of \code{spvspv} is captured via the
following hyper-triple:
\begin{equation}
  \label{eq:spvspv}
  {\small{
  \hspec{\!\!\!
  \begin{array}{l}
    \arrs_{\texttt{y}}(\cdot) \str \\[2pt]
    \arrs_{\texttt{x}}(\cdot)
  \end{array}
  \!\!}    
  \hmid{\!\!\!
    \begin{array}{c@{\ }c@{\ }l}
    1 &:& \mathtt{spvspv(y_{ind}, y_{val}, x_{ind}, x_{val})}
    \\ [2pt]
    \angled{2, j}_{j\in [0,N)} &:&  \mathtt{sv\_get(y_{ind}, y_{val},} j)
    \\ [2pt]
    \angled{3, j}_{j\in [0,N)} &:& \mathtt{sv\_get(x_{ind}, x_{val},} j)
    \end{array}
  \!}
  \hspec{\!\!\!
  \left.
  \begin{array}{r}
  s \\[2pt] \many{y} \\[2pt] \many{x}
  \end{array}
  \right|
  \begin{array}{l}
    \pure{s= \sum_j \many{y}(j)\many{x}(j)}
  \end{array}
  \!\!}
  }}
\end{equation}
The predicate $\arrs_{\texttt{y}}(\cdot)$ abbreviates
{\small{\( \arr{y_{ind}(\cdot)}{y_{ind}} \str \arr{y_{val}(\cdot)}{y_{val}} \)}},
similarly for $\arrs_{\texttt{x}}(\cdot)$.
To see how can we get~\eqref{eq:spvspv_dom} out of~\eqref{eq:spvspv}
recall that in~\eqref{eq:spvspv_dom}, the implicitly universally
quantified (outside of the entire hyper-triple) variable $i$ ranges
from $0$ up to $|\inds{m}| - 1$.
Since $i$ is \emph{fixed} across all programs in the second component,
we get rid of it via \emph{domain substitution}.

\begin{wrapfigure}[7]{r}{0.46\textwidth}
  \setlength{\abovecaptionskip}{10pt}
  \vspace{-15pt}
  \begin{minipage}{0.9\linewidth}
  {\small{  
  \centering
  \begin{mathpar}
    \inferrule[\tname{Subst}]
    {
      \begin{array}{c}
        \phi : S \to S' $ is injective$ \quad
        \forall i, \pp'(i) = \pp(\phi(i)) \arcr[2pt]
        \hslcc{ \phi[P] }{\phi[S]:\ \pp'}{ \phi[Q] }
      \end{array}
    }
    {
      \hslcc{ P }{S:\ \pp}{ Q \!\ }
    }
  \end{mathpar}
  }}
  \end{minipage}
  \caption{Domain substitution rule}
  \label{fig:subst}
\end{wrapfigure}
A simplified version of the substitution rule \tname{Subst} is
presented in~\autoref{fig:subst}. 
Its key component is an \emph{injective} mapping $\phi$ from the index
set $S$ to a set $S'$, which embodies the reindexing in question. This
rule applies~$\phi$ to the index set of the triple $S$. It also
changes $\pp$ to $\pp'$, which is a reindexed $\pp$ \wrt to $\phi$.
Finally, it re-indexes the components of both pre- and postcondition
by applying $\phi$ (denoted, \eg, $\phi[P]$), transforming, in
particular, spatial assertions of the kind $x(\iota) \mapsto v$ into
$x(\phi(\iota)) \mapsto v$. For composite assertions, containing separation conjunction, and big separation conjunction, $\phi$ just iteratively distributes over their logical connectives.
This transformation does not affect the validity of the assertions,
since $\phi$ is injective.
%

We apply \tname{Subst} to~\eqref{eq:spvspv_dom}, taking $S$ to be the
index set of~\eqref{eq:spvspv_dom}, and $\phi$ to be a function
removing the second element $i$ from all tuples from the second
component of $S$. Note that this $\phi$ is going to injective as~$i$
is fixed across all the programs in the second component. 
The obtained triple is:
\begin{equation}
  \label{eq:spvspv-m}
  {\small{
  \hspec{\!\!\!
  \begin{array}{l}
    \arrs(\cdot) \\[2pt]
  \end{array}
  \!\!}    
  \hmid{\!\!\!
    \begin{array}{c@{\ }c@{\ }l}
    1 &:& \mathtt{spvspv(m_{ind}[i], m_{val}[i], x_{ind}, x_{val})}
    \\ [2pt]
    \angled{2, j}_{j\in [0,N)} &:&  \mathtt{sv\_get(m_{ind}[i], m_{val}[i],} j)
    \\ [2pt]
    \angled{3, j}_{j\in [0,N)} &:& \mathtt{sv\_get(x_{ind}, x_{val},} j)
    \end{array}
  \!}
  \hspec{\!\!\!
  \left.
  \begin{array}{r}
  s \\[2pt] \many{m} \\[2pt] \many{x}
  \end{array}
  \right|
  \begin{array}{l}
    \pure{s= \sum_j \many{m}(j)\many{x}(j)}
  \end{array}
  \!\!}
  }}
\end{equation}

\noindent
Now, substituting $\small\tinds{m}[\mathtt{i}]$ for \code{y}
in~\eqref{eq:spvspv}, we get a hyper-triple that
immediately implies~\eqref{eq:spvspv-m}.


\begin{samepage}

\subsection{A Rule for While-Loops}
\label{sec:while}
%
%
What is left now is to verify correctness of \code{spvspv} stated as a
hyper-triple~\eqref{eq:spvspv}.
Below, we sketch its proof, primarily focusing on how to deal with the
\code{while}-loop in the body of \code{spvspv}.

\begin{wrapfigure}[8]{r}{0.35\textwidth}
  \setlength{\abovecaptionskip}{7pt}
  \begin{minipage}{0.9\linewidth}
    \begin{center}
      \includegraphics[width=1\linewidth]{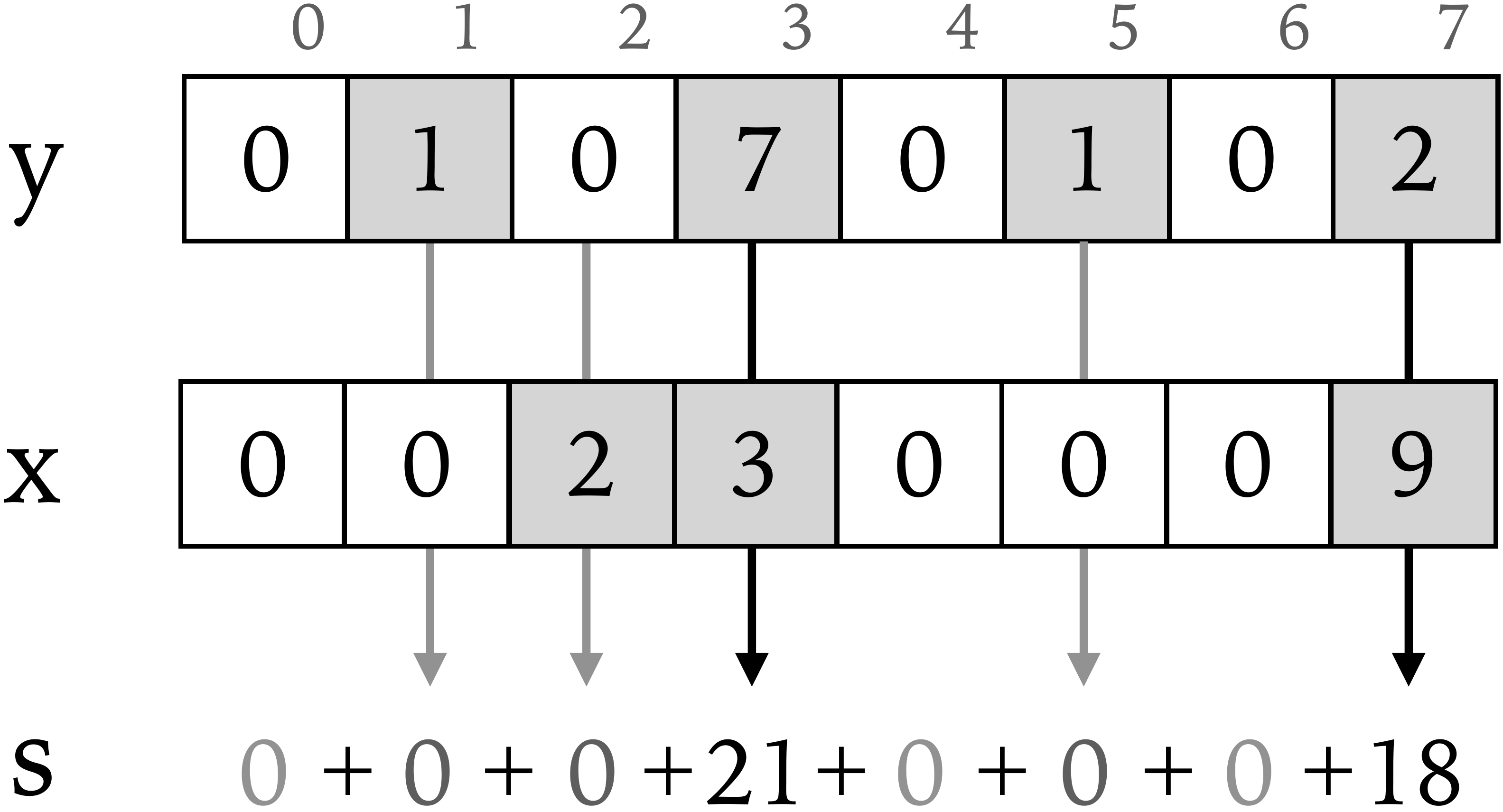}
    \end{center}
  \end{minipage}
\caption{SpVSpV pictographically}
\label{fig:spvspv}
\end{wrapfigure}
The proof is enabled by an important fact: the \code{while}-loop in
\code{spvspv} processes \emph{only non-zero} values of input vectors.
An illustration of how \code{spvspv} works is depicted
in~\autoref{fig:spvspv}.
Given the vectors $y$ and $x$, the algorithm visits all positions
where at least one vector has a non-zero value (1, 2, 3, 5, and 7 in
this example).
For each such position, \code{spvspv} will increase the value of
\code{s} only if \emph{both} positions are non-zero (such situations
are depicted with solid black lines in~\autoref{fig:spvspv}). Getting
back to~\eqref{eq:spvspv}, this suggests us to align each iteration of
the \code{while}-loop with the corresponding element from the sequence
of non-zero indexes in payloads of either of the two vectors
$\inds{y}\cup \inds{x}$.
For all other indices $j$, results of both \code{sv_get} functions,
$\many{y}(j)$ and $\many{x}(j)$ will be zero, hence they can be safely
elided from the summation $\sum_j \many{y}(j)\many{x}(j)$. This can be
captured by the following triple:

\end{samepage}

\begin{equation}
  \label{eq:spvspv_zero}
  {\small{
  \hspec{\!\!\!
  \begin{array}{l}
    \arrs_{\texttt{y}}(\cdot) \str \\[2pt]
    \arrs_{\texttt{x}}(\cdot)
  \end{array}
  \!\!}    
  \hmid{\!\!\!
    \begin{array}{c@{\ }c@{\ }l}
    \angled{2, j}_{j\not\in \inds{y} \cup \inds{x}} &:&  \mathtt{sv\_get(y_{ind}, y_{val},} j)
    \\ [2pt]
    \angled{3, j}_{j \not\in \inds{y} \cup \inds{x}} &:& \mathtt{sv\_get(x_{ind}, x_{val},} j)
    \end{array}
  \!}
  \hspec{\!\!\!
  \left.
  \begin{array}{r}
  \many{y} \\[2pt] \many{x}
  \end{array}
  \right|
  \begin{array}{l}
    \pure{\forall j \not\in \inds{y} \cup \inds{x}, \many{y}(j) = 0} \str \\[2pt]
    \pure{\forall j \not\in \inds{y} \cup \inds{x}, \many{x}(j) = 0}
  \end{array}
  \!\!}
  }}
\end{equation}

\noindent
The hyper-triple~\eqref{eq:spvspv_zero} can proven in a way similar to
the specification~\eqref{eq:spmspv4} of \code{ucsr_get}.
First, using the \tname{Product} rule we reduce it to a set of unary
triples, and then, noticing that the \code{index} function in the body
of \code{sv_get} will return~\code{-1} for any $j$ that is not an
element of the set $\inds{x} \cup \inds{y}$.
Now applying the \tname{Focus} rule, and taking
$\ind \eqdef \inds{y} \cup \inds{x}$, we reduce~\eqref{eq:spvspv} to
the following goal:
\begin{equation}
  \label{eq:spvspv_frame}
  {\small{
  \hspec{\!\!\!
  \begin{array}{l}
    \arrs_{\texttt{y}}(\cdot) \str \\[2pt]
    \arrs_{\texttt{x}}(\cdot)
  \end{array}
  \!\!}    
  \hmid{\!\!\!
    \begin{array}{c@{\ }c@{\ }l}
    1 &:& \mathtt{spvspv(y_{ind}, y_{val}, x_{ind}, x_{val})}
    \\ [2pt]
    \angled{2, j}_{j\in \ind} &:&  \mathtt{sv\_get(y_{ind}, y_{val},} j)
    \\ [2pt]
    \angled{3, j}_{j\in \ind} &:& \mathtt{sv\_get(x_{ind}, x_{val},} j)
    \end{array}
  \!}
  \hspec{\!\!\!
  \left.
  \begin{array}{r}
  s \\[2pt] \many{y} \\[2pt] \many{x}
  \end{array}
  \right|
  \begin{array}{l}
    \pure{s= \sum\limits_{j \in \ind} \many{y}(j)\many{x}(j)}
  \end{array}
  \!\!}
  }}
\end{equation}

\begin{figure}
  \setlength{\abovecaptionskip}{5pt}
  \setlength{\belowcaptionskip}{-10pt}
    \begin{minipage}{1.0\linewidth}
    {\small{
    \begin{mathpar}
    \!\!\!\!
    \inferrule[\tname{While}]
    {
    \begin{array}{c}
    S = \bigcup_{i=0}^{N - 1} S_i
    \quad
    \local{R_i}{S_i}
    \quad
    \begin{array}{l}
      \forall \many{x}, \hsllazy{I(N,\many{x})\ \ }{\iota: c\ \ }{b}{\pure{b=\mathit{false}}} \arcr[2pt]
      \forall \many{x}, i < N,
      \hsllazy{
        I(i, \many{x}) \str
        R_i \ \ }
        {\iota : \ifc{c}{p},~~S_i : \pp_i}
        {z, \many{y}}
        {I(i+1, \many{x}\many{y})}
    \end{array}
    \end{array}
    }
    {
    \hsllazy{
            I(0, \many{\emptyset}) \str
            \bigstr_{i=0}^{N - 1} R_i \ \ }
            {\iota : \whilel{c}{p},\
             S : \pp}
            {z, \many{y}}
            {I(N, \many{y}) }}
    \end{mathpar}
    }}
    \end{minipage}
    \caption{A rule for \kw{while}-iteration.}
    \label{fig:while}
\end{figure}

\noindent
Finally, we are ready to align each iteration of the \code{while}-loop
with the corresponding element of $\ind$ by applying the \tname{While}
rule shown on~\autoref{fig:while}.
\tname{While} is (predictably) similar to the \tname{For} rule. 
The intuition for this rule comes from the fact that each loop
$\whilel{c}{p}$ that terminates after no more than $N$ steps is
equivalent to just $N$ iterations of $\ifc{c}{p}$. 
Therefore, if we divide our index set into $N$ batches, we can align
each batch with one execution of $\ifc{c}{p}$. Note that this will
also handle the case when the loop terminates after $K < N$ steps: in
that case all runs of $\ifc{c}{p}$ will become idle, starting from the
$K$\ith step. 
A new premise on the top right corner will ensure that the loop will
terminate once we have exhausted all batches.\footnote{\logic is
  implemented as logic for \emph{total} correctness, yet the
  \rulename{While} rule does not require an explicit termination
  measure. The termination will follow from the fact that, each
  iteration is aligned with its own batch from a finite set of index
  batches.}
To proceed with the proof of~\eqref{eq:spvspv_frame} by making use of
the \tname{While} rule, we divide our index set as follows:

\vspace{-5pt}

{\small{\[ \angled{2,j}_{j \in \ind} \cup \angled{3,j}_{j\in\ind} = 
  \bigcup_{j=0}^{|\ind|-1} \set{\angled{2, \ind[j]}, \angled{3,
      \ind[j]}} \]}}

\vspace{-3pt}

\noindent
Here, to reference to the $j$\ith component of $\ind$ we treat as an sorted sequence without duplicates. 
Let us show the most important part of the loop invariant and sketch
the rest of the prove:
\noindent
\begin{equation} 
  \label{eq:csr8}
    \small{
    \begin{array}{r@{\ }c@{\ }l@{\ }l}
      \Ispvspv(j,\many{y}\many{x}) & \eqdef 
     & \exists \Iyy, \Ixx, & \mcode{iY} \mapsto\Iyy \str \mcode{iX}
                             \mapsto \Ixx \str \\[2pt]
     &&& s \mapsto \sum_{k < j} \many{y}(\ind[k])\,\many{x}(\ind[k]) \str  \\[2pt]
     &&& 
\pure{\min(\inds{x}[\Ixx], \inds{y}[\Iyy]) = \ind[j]} \str\dots
    \end{array}}
  \end{equation}
The first line of the invariant simply gives names to the values
stored in the vector counters ($\Iyy$~and~$\Ixx$).
The second line captures the fact that $s$ stores the accumulated
intermediate result of the dot product.
The third line is the most subtle:
it states that at each iteration, the current position in the $\ind$
sequence, is equal to a \emph{minimum} of the current positions in
both vectors ($\inds{y}[\Iyy]$ and $\inds{x}[\Ixx]$). 
%
%
If $\inds{y}[\Iyy]$ and $\inds{x}[\Ixx]$ are same, then they are both equal to $\ind[k]$. Hence line~6 of \code{spvspv} will increase the sum by $\many{y}(\ind[j])\, \many{x}(\ind[j])$, preserving the second line of $\Ispvspv$.
In other cases, using the third line of the invariant, we will be able
to show one of $\inds{y}[\Iyy]$ or $\inds{x}[\Ixx]$ to be zero, so the
sum will indeed remain unchanged, matching the code on lines~8-9
and~10-11 of \autoref{fig:smpv}.
%
%

  
\subsection{Putting It All Together}

And that concludes our tour of \logic!
As a quick recap, the right side of \autoref{fig:mech} summarises the
main milestones of our proof of \code{spmspv}'s
specification~\eqref{eq:spmspv1} (with pre-/posts elided), showing the
corresponding sections and changes in the proof goals.
Once again, we emphasise the key \logic aspects that enabled the
proof: \emph{separation} for modular reasoning (\autoref{sec:prod}),
\emph{parametrisation} for deconstructing loops
(\autoref{sec:for}~and~\ref{sec:while}), and \emph{compositionality}
for proof reuse (\autoref{sec:subst-spec}).
We next briefly outline \logic's soundness argument, postponing the
presentation of its implementation as a verification tool till
\autoref{sec:mech}, in which we will show the Coq mechanisation of our
\code{spmspv} example.


\section{\logic, Formally}
\label{sec:logic}

%
The programming language of \logic (\autoref{fig:prog}) is
chosen to be generic enough to model tensor computations in C, Python,
and Julia.
Its commands include location and array access ($!x$ and $x[e]$),
allocation/deallocation for both individual memory locations (via
{\small{\kw{alloc}/\kw{free}}}) and arrays (via {\small{\kw{malloc}/\kw{mfree}}}).
The {\small\kw{let}}-expressions are also used to chain sequences of
commands, thus encoding a sequential composition.
%
%
%
For the sake of the presentation, we assume that all the considered
programs are well-typed, integers are unbounded, and division is not
present amongst the operations.

\begin{figure}[t]
\setlength{\abovecaptionskip}{5pt}
\setlength{\belowcaptionskip}{-5pt}
\[\footnotesize\begin{array}{l@{\ }c@{\ }l}
  \text{Locations}         & x,\ y,\ \dots & \text{alpha-numeric strings} \\
  \text{Constants}         & n     & \text{unbounded integer literals} \\
  \text{Expressions}       & e ::= & n~|~\True~|~\False~|~x~|~e = e~|~e < e~|~e \land
                                     e~|~\neg e~|~e~\oplus~e \\
  \text{Command}           & c ::= & !x~|~x[d]~|~x := e~|~x[e] := e~|\
                                     \forl{x}{e}{e}{c}~|~\kw{while}~(e)~{c}~|~\kw{if}~(e)~\{c\}~\kw{else} \{c\}~| \\
&& \kw{let}~x := c~\kw{in}~c~|~\kw{alloc}(n)~|~\kw{malloc}(n)~|~\kw{free}(x)~|~\kw{mfree}(x)~|~\kw{length}(e)~|~\kw{return}~e
%
\end{array}\]
\caption{\logic programming language}
\label{fig:prog}
\end{figure}

\subsection{Semantics}

We adopt a standard big-step semantics for our language. 
Let $c$ be a command and $s$ be a \emph{state}, \ie, a finite mapping
from locations $\Loc$ to values $\Val$. The evaluation relation
$\angled{c,s} \Downarrow \angled{v, s'}$ is defined inductively on the
structure of $c$. 
It means that a command $c$ starting from input store $s$ terminates
with the return value $v \in \Val$, and store $s'$.%
\footnote{We only consider terminating programs, and require
  termination measure given for programs with \fcode{while}-loops.}
In this work, we only focus on deterministic program executions, \ie,
such that allocation does not exercise randomness.
%
%
%
For a fixed index set $S$, given two \emph{hyper-heaps} $h$ and $h'$
defined as mappings from $\Loc \times S$ to values, a hyper-value
$\many{v}$ and an $S$-indexed program product~$\pp$, the semantics of
a product execution is defined as follows:
\begin{equation*}
{\small{
\angled{\pp,h} \Downarrow_S \angled{\many{v}, h'} \eqdef \forall \iota \in S, \angled{\pp(\iota),h|_\iota} \Downarrow \angled{\many{v}(\iota), h'|_\iota}
}}  
\end{equation*}
That is, for every $\iota \in S$ and individual program $\pp(\iota)$
executed on the corresponding projection $h|_\iota$ of the initial
hyper-heap (defined as
$h|_\iota \eqdef \lambda l \in \Loc.\ h(l, \iota)$), its execution
results in a new state $h'|_\iota$ and a value $\many{v}(\iota)$.
Similarly to the ordinary (unary) programs, we assume all programs in
a product to be deterministic and terminate (the latter is guaranteed
for all programs that verify in \logic).
The details of the semantics can be found in our Coq mechanisation~\cite{lgtm}.




%
Hyper-safety triples in \logic are defined in terms of the weakest
precondition predicate:
\begin{equation}
\label{eq:wp}
{\small{
\wp{S:\ \pp}{Q}\ \eqdef\ \lambda h,\ \exists h'\ \many{v},\ \angled{\pp,h} \Downarrow_S \angled{\many{v}, h'} \land Q(\many{v})(h')    
}}
\end{equation}
The weakest preconditions are relative to a provided postcondition
(\ie, a predicate on hyper-heaps)~$Q$ and are defined as predicates on
input heaps~$h$, where $Q$ is a function from hyper-value to
hyper-heap assertions.
%
%
We next define a hyper-triple $\hslcc{P}{S: \pp}{Q}$ as a notation for
$P \vdash \wp{S:\ \pp}{Q}$.
In plain language, this means that an assertion $P(h)$ implies
$\wp{S:\ \pp}{Q}(h)$ for any hyper-heap $h$. 
We will use the expanded version of the postcondition whenever we want
to emphasise its assertion $Q$ to be a function of the result hyper-value:
$\hslc{H}{S:\pp}{\many{v}}{Q(\many{v})}$ and we will omit the index set~$S$ whenever it
is clear from the context. 
Note that, due to the termination requirement on the programs in
products, all triples in \logic ensure total correctness by
definition.


\subsection{Support for Reasoning with Parametrisation and Separation}
\label{sec:kill-lhc}

Many essential rules of \logic take their origins in the Logic for
Hyper-triple Composition (LHC) by~\citet{DOsualdoFD22}. \logic's
\emph{structural} and \emph{lockstep} rules directly adjust those of
LHC to work with parametrised triples that use Separation Logic (SL)
connectives and predicates.
Below, we outline the key \logic's enhancements on top of LHC,
supporting parametrisation and separation.

\subsubsection{Parametrisation}

\begin{wrapfigure}[5]{r}{0.42\textwidth}
\setlength{\abovecaptionskip}{4pt}
\vspace{-18pt}
\begin{minipage}{1\linewidth}
{\small{
  \begin{prooftree}
    \AXC{idx\((Q_1) \subseteq \text{supp}(t_1)\ \) idx\((Q_2) \subseteq \text{supp}(t_2)\)}
    \noLine
    \UIC{\(\mathsf{wp}\ \pp_1 \{ Q_1 \} \land \mathsf{wp}\ \pp_2 \{ Q_2 \}\)}
    \RightLabel{\small \tname{Wp-Conj}}
    \UIC{\( \mathsf{wp}\ (\pp_1 + \pp_2)\ \{Q_1 \land Q_2 \} \)}
  \end{prooftree}
}}
\end{minipage}
\caption{\tname{WpConj} rule from LHC} 
\label{fig:wpconj} 
\end{wrapfigure}

One example of an \logic rule that is not derivable in LHC is the
\tname{Product} rule (\autoref{fig:seq}). 
\tname{Product} takes inspiration from LHC \tname{WpConj} rule,
depicted on~\autoref{fig:wpconj}.
LHC triples do not feature the index set explicitly, as all programs
in LHC triples can only be indexed with natural numbers.
This rule allows one to conjoin specifications of two program products
$\pp_1$ and $\pp_2$ with possibly overlapping components, yielding
$\pp_1 + \pp_2$, if $\pp_1$ and $\pp_2$ agree on the overlapping part.
Even though LHC's rules are phrased in terms of triples with an
arbitrary number of programs, this number must be \emph{fixed} prior
the proof.
%
%
This is not a problem in LHC: for program products of a fixed known
size $n$, one would have to apply this rule $n$ times.


Things become problematic with parametrised triples: to get the effect
of \logic's \tname{Product} rule, one would have to apply
\tname{Wp-Conj} \emph{the size of $S$} number of times.
To support cases when $|S|$ depends, \eg, on a program parameter, this
would require LHC support a ``constant-space'' iterated rule
application (as in $\mathcal{R}^n()$ instead of
$\mathcal{R}(\mathcal{R}(\dots))$ - repeated explicitly $n$ times) of
\tname{Wp-Conj}.
This ability does, indeed, come ``for free'' for an embedding into a
higher-order logic such as one of Coq (hence the importance of Coq
formalisation in our work). 
It is, however, missing from LHC, but is explicitly captured in LGTM
rules. 
%
%
%
The same idea applies to \tname{For} and \tname{While} rules
(\autoref{sec:for}--\ref{sec:while}). In both those rules we divide
the index set $S$ into $n$ batches, where $n$ is a parameter coming
from the specification, so that one iteration of the
\kw{for}-/\kw{while}-loop handles only one such batch.

\subsubsection{Separation}
\label{sec:logic-sep}

\begin{wrapfigure}[3]{r}{0.38\textwidth}
  \setlength{\abovecaptionskip}{7pt}
  \vspace{-22pt}
  \begin{minipage}{1\linewidth}
  {\small{
$ \hslc{\tt
          emp}{S:\kw{alloc}(v)}{\many{x}}{\bigstr\limits_{i\in
            S}\many{x}(i)\mapsto v} $
  }}
  \end{minipage}
  \caption{\logic rule for allocation} 
  \label{fig:alloc} 
\end{wrapfigure}

\logic extends LHC's \emph{lockstep} inference rules with Separation
Logic specific-rules (\eg, for pointer and array allocation). 
Adding required adjusting a unary SL proof system to
work with hyper-heaps, as, for instance, is exemplified by
\tname{Alloc} rule presented on~\autoref{fig:alloc}. 
\tname{Alloc} states that the results of a family of pointer
allocations are these pointers $\many{x}$, pointing to the assigned
value. Other \logic lockstep rules follow the same intuition and are
available in \autoref{sec:appendix}.

\subsection{Soundness and Mechanisation}
\label{sec:soundness}

We say that \logic triple $\hslcc{P}{S:\pp}{Q}$ is \emph{derivable}
and write $\vdash_{\sf\logic} \hslcc{P}{S:\pp}{Q}$ if this triple can
be obtained using the rules of the logic.
We say that the triple $\hslcc{P}{S:\pp}{Q}$ is \emph{semantically
  valid} and write $\models \hslcc{P}{S:\pp}{Q}$ if and only if
\begin{equation*}  
{\small{
\forall h, P(h) \Rightarrow \exists h'\
      \many{v}, \angled{\pp, h} \Downarrow_{S} \angled{h',\many{v}}
      \land Q(\many{v})(h')
}} 
\end{equation*}
That is, we ask that for every hyper-heap $h$ satisfying the
precondition, if we run $\pp$ on it, it terminates, and the resulting
hyper-heap and hyper-value satisfy $Q$.
The derivability of a logic triple in \logic is connected to its
semantic validity by the following standard soundness result:

\begin{theorem}[Soundness]
\label{thm:soundness}
  For all hyper-heap assertions $P$, functions from hyper-value to
  hyper-heap assertions $Q$, and program products $\pp$ indexed by a set
  $S$,
  \[\vdash_{\sf{\logic}} \hslcc{P}{S:\pp}{Q} \Rightarrow~~\models \hslcc{P}{S:\pp}{Q}\]
\end{theorem}
We have mechanised the meta-theory of \logic and its semantics in the
logic of Coq proof assistant. 
In our mechanisation, we defined the Hoare triples semantically in
terms of the weakest precondition predicates, which, by their
definition~\eqref{eq:wp} encode semantic validity.
We then defined each \logic rule as a lemma and proved that it holds,
thus establishing the result of \autoref{thm:soundness}.
Our mechanisation of \logic is built by extending
CFML~\cite{Chargueraud11,Chargueraud20}---a mechanised sequential
Separation Logic.
Our changes to the original CFML including the theory of hyper-heaps,
the \textsf{wp}-calculus for program products, \logic rules, and
automation, are totalling at 20kLOC.
%

\section{\logic as a Verification Tool}
\label{sec:mech}

Let us now showcase \logic as a verification framework by walking
through a mechanisation of the \code{spmspv} case study from
\autoref{sec:overview} and explaining its most essential proof
automation components.


\subsection{Structuring Mechanised Proofs}

\autoref{fig:mech} presents a mechanised proof of \code{spmspv}
(left) next to the informal outline of the proof
derivation from \autoref{sec:overview} (right), with gray rectangles
depicting the transformations of the proof goals.

\begin{figure}[t]
\setlength{\abovecaptionskip}{7pt}
\setlength{\belowcaptionskip}{-10pt}
  \includegraphics[width=\linewidth]{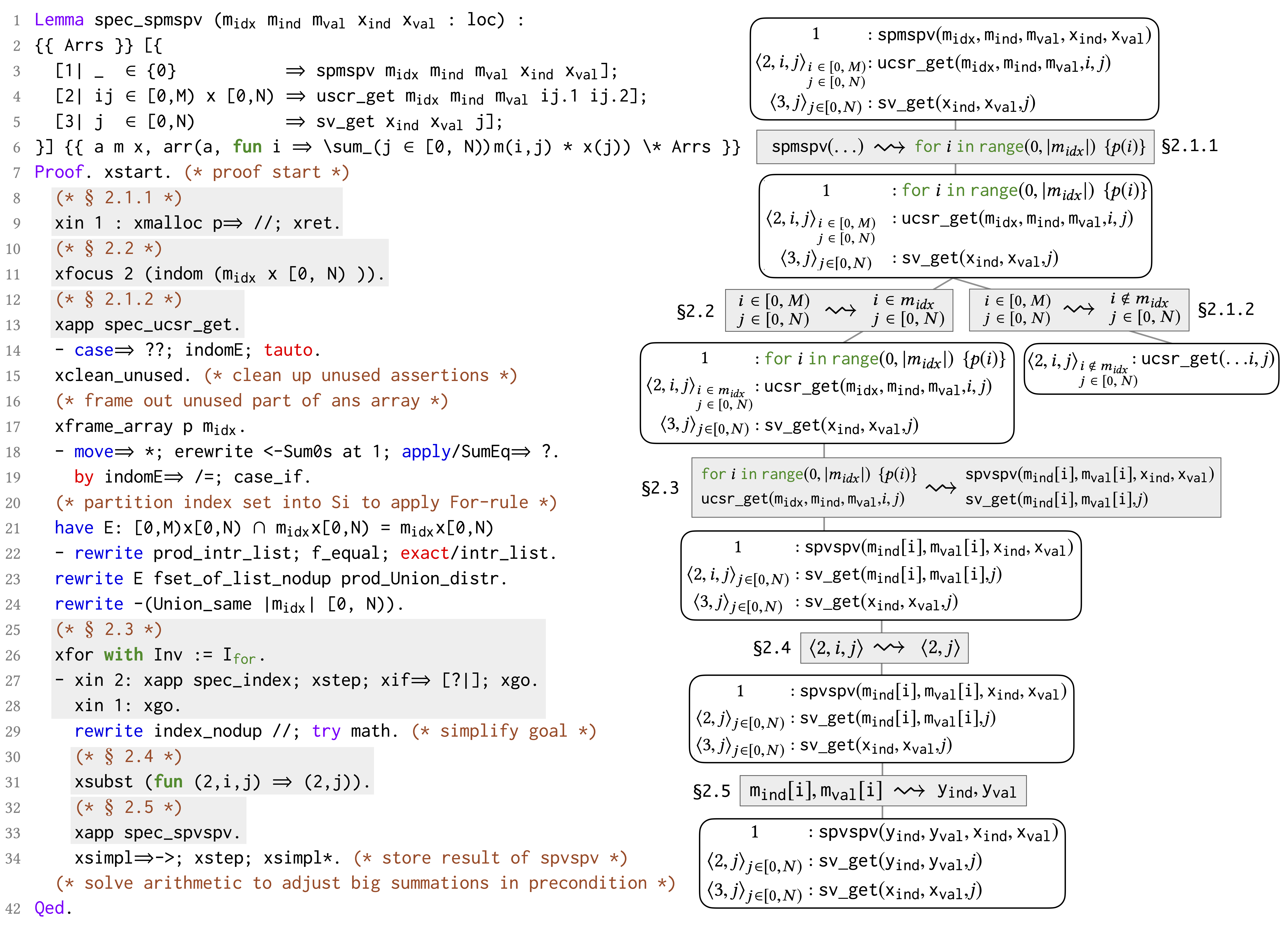}
  \caption{Mechanised \logic proof of \code{spmspv} (left) and the corresponding paper-and-pencil derivation (right).}
  \label{fig:mech}
\end{figure}

One noticeable technical difference between the two proofs is the
representation of index sets in the mechanisation.
In the Coq specification (lines~1-7), each component of the program
product has an explicit \emph{tag}, separate from the rest of the
index set definition.
For example, the tag of \code{uscr_get} is \code{2}.
We tried several designs for \logic triples and found the ``tagged
union'' of index sets to be the most user-friendly when it comes to
the local per-component reasoning via focusing
(\autoref{sec:nest-frame}).\footnote{The tag-based encoding does not
  affect expressivity, since index sets can be arbitrarily mapped
  using \rulename{Subst} (\autoref{fig:subst}).}

The proof immediately makes use of the index set tags: the \code{xin}
\emph{tactical} (\ie, a higher-order tactic) at line~9 takes the
tag~\code{1} as its argument, thus applying a version of the
\rulename{SeqU1} rule (\autoref{sec:seq}).
Using \code{xin}, one can advance the proof of a composite
hyper-triple by symbolically executing instructions in its
sub-components in a lockstep manner by passing a sequence of tactics
implementing sequential (or, in general, lockstep) rules after the
\code{:} separator.
In this case, we pass two tactics to the \code{xin} tactical. The
first one, \code{xmalloc}, symbolically executes the \kw{malloc}
instruction in the first component. The second, \code{xret}, strips
off the last \kw{return} statement in the body of \code{spmspv}.

Moving on with the proof, we achieve the effect of the \tname{Focus}
rule from~\autoref{sec:nest-frame} by calling \code{xfocus} tactic at
line~11. 
As the result, the index set of the second component will be divided
into two groups, as per the tactic's second argument: inside
$\idxs{m}\times[0,N)$ and outside of it.
Subsequently, the lines~12-13 verify the runs of \code{uscr_get}
outside of $\idxs{m}\times[0,N)$, using the specification of
\code{uscr_get} via \code{xapp} tactic.
In the rest of the proof, we first divide the index set into a union
of $|\idxs{m}|$ batches, to apply the \tname{For} rule (lines~21-24).
Next, at line~26 we apply the \code{xfor} tactic, replicating the
steps from~\autoref{sec:for}, followed by an application of the
\code{xsubst}, which performs the domain substitution
from~\autoref{sec:subst-spec}.
We conclude the proof by using an independently verified specification
of \code{spvspv} (line~33), discharging the residual obligations about
index set manipulations (lines~34-42).


\vspace{-5pt}

\subsection{Proof Automation for Separation and Parametrisation}

The two enabling features of \logic, modularity via separation and
parametrisation, come with additional obligations in rules that
exercise them. Those obligations have to do with (a)~proving
\emph{assertion locality} (\autoref{sec:prod}) when reducing triple
arity and (b)~\emph{changing indices within assertions} when
performing domain substitution with arbitrary index sets
(\autoref{sec:subst-spec}).
Below, we describe two proof automation techniques that eliminate the
proof burden associated with those obligations.

\vspace{-3pt}

\subsubsection{Discharging Locality Obligations}

Tactics that generate proof obligations for locality of hyper-heap
assertions
include \code{xprod} (which implements the \tname{Product} rule) and
\code{xfor} (which implements \tname{For}). 
For instance, the application of \code{xfor} at line~26 of
\autoref{fig:mech} generates a number of subgoals requiring one to
establish locality of the assertions~$R_i$ used as preconditions for
programs in the respective components (\cf~\autoref{fig:forsimpl}).
Even a subgoal for the third component looks a bit intimidating:

%
%

\vspace{-4pt}

{\small{\[
  \code{local}\left(\left(
    \bigstr_{i = 0}^{N-1} \arrl{\tinds{x}\angled{3,i}}{\inds{x}} \str 
    \bigstr_{i = 0}^{N-1} \arrl{\tvals{x}\angled{3,i}}{\vals{x}}\right),\ 
\set{\angled{3,i} : i \in[0,N)} \right)
  \]}}

\vspace{-1pt}

\begin{wrapfigure}[10]{r}{0.37\textwidth}
  \setlength{\abovecaptionskip}{5pt}
  \vspace{-17pt}
  \begin{minipage}{0.9\linewidth}
  {\small{ 
\begin{lstlisting}[language=Coq,xleftmargin=10pt,xrightmargin=0pt,basicstyle=\scriptsize\ttfamily,numbersep=5pt]
Ltac xlocal := intros;
  match goal with 
  | |- local (P1 \* P2) _ => 
       apply local_conj; xlocal
  | |- local (\*_(i \in s) Pi) _ => 
       apply local_big_conj; xlocal
  | |- local (arr(p(i), l)) s => 
       apply local_array; indomE
  ...
  end
\end{lstlisting}
}}
  \end{minipage}
  \caption{\texttt{xlocal} tactic}
  \label{fig:xlocal}
\end{wrapfigure}
To dispatch such goals, we provide the \code{xlocal} tactic
shown~\autoref{fig:xlocal}.
In a nutshell, this tactic, structurally decomposes the hyper-heap
assertion in a goal, reducing locality of a composite assertion to
locality of its parts. For instance, lines~3-4 of the tactic
reduce locality of $P_1 \str P_2$ to locality of both $P_1$
and $P_2$, calling \code{xlocal} recursively on those subgoals. 
Lines~5-6 we reduce locality of $\bigstr_{i\in S} H_i$ to the
locality of $P_i$ for each $i \in S$. 
Finally, whenever \code{xlocal} reaches terminal goals such as
$\code{local}(\code{arr}(p(i),l), S)$, it applies the
\code{local_array} lemma, resulting in the obligation $i \in S$,
proven by the \code{indomE} membership solver from the MathComp
library~\cite{Maboubi-Tassi:MathComp}.

\vspace{-3pt}

\subsubsection{Automating Index Set Substitution}
\label{sec:auto}
%
%
The need for proof automation in the presence of parametrisation
manifests whenever we want to perform reindexing involving arbitrary
index sets via the \tname{Subst} rule.
As \tname{Subst} applies a substitution function $\phi$ to both pre-
and postcondition, $\phi$ must be distributed over the connectives of
the \logic including, separation conjunction and its iterated version
(big-$\bigstr$). One na\"{i}ve approach to implement such distribution
automatically would involve proving a lemma of the form
$\phi\left[\bigstr_{i\in S} H_i\right] = \bigstr_{i\in S}
\phi\left[H_i\right]$ for distributing the substitution over
$\bigstr$, and use it for repeated rewriting in the assertions.
Unfortunately, this solution does not combine well with the shallow
embedding of ``big'' operators into Coq, that implement indexing
over~$i$ via Coq's lambda-functions.
As the inner assertions $H_i$ of $\bigstr_{i\in S} H_i$ might depend
on the lambda-bound~$i$, rewriting them would require independently
instantiating the equality lemmas such as
$\forall P, Q.~\phi[P{\str}Q] = \phi[P]{\str}\phi[Q]$ with assertions
that depend on~$i$ and are not well-formed outside of the lambda.
Therefore, after we rewrite an assertion involving $\bigstr$ using the
above equality once, the application of $\phi$ will occur in
the scope of the binder~$i$, preventing its further rewrites.

\begin{wrapfigure}[10]{r}{0.40\textwidth}
  \setlength{\abovecaptionskip}{0pt}
  \vspace{-15pt}
  \begin{minipage}{1\linewidth}
  {\small{ 
    \begin{tabular}{c}
\begin{lstlisting}[language=Coq,xleftmargin=0pt,xrightmargin=0pt,basicstyle=\scriptsize\ttfamily,numbersep=5pt]
Ltac xsubst := match goal with 
  | |- \phi[P1 \* P2] = ?H => 
       rewrite subst_conj; 
       apply conj_eq; xsubst
  | |- \phi[\*_(i \in s) Pi] = ?H => 
       rewrite subst_big_conj; 
       apply big_conj_eq; intros; xsubst
  | |- \phi[arr(p(i), l)] = ?H => 
       rewrite subst_array; reflexivity
  ...
  end
\end{lstlisting}
\end{tabular}
}}
  \end{minipage}
  \caption{\texttt{xsubst} tactic}
  \label{fig:xlocal}
\end{wrapfigure}
We overcome this hurdle with the following trick.
When performing domain substitution via~$\phi$, we start with
replacing the precondition $\phi$\code{[P]}
(\emph{resp.}~$\phi$\code{[Q]}) with a fresh existential variable
\qmark\code{H}, adding a subgoal $\phi$\code{[P]} = \qmark\code{H} to the
set of proof obligations.
This subgoal will be repeatedly transformed as we gradually propagate
$\phi$ inside \code{P}, elaborating \qmark\code{H} as we go.
This idea is implemented by \code{xsubst} tactic shown in
\autoref{fig:xlocal}. 
To understand its effect on the proof, consider the lines~6-8 of its
definition that deal with the big-$\bigstr$ operator.
The tactic implementation first propagates $\phi$ under the big-star
by performing the rewrite via the \code{subst_big_conj} lemma,
followed by application of the \code{big_conj_eq} lemma of type
$\left(\forall i, H_i = P_i\right) \Rightarrow \bigstr_{i\in S} H_i =
\bigstr_{i\in S} P_i$, which is essentially a specialised version of
the functional extensionality axiom.
Applying this lemma instantiates the existential variable \qmark\code{H}
with $\bigstr_{i =0}(\qmark\mcode{Hi}~i)$, producing a subgoal
$\forall i, (\phi[P_i] = \qmark\mcode{Hi}~i)$, which will require further
elaboration of \qmark\code{Hi}.
The tactic subsequently fixes the binder $i$ by moving it to the proof
context using \code{intros}, followed by a recursive call of
\code{xsubst}.
%

When \code{xsubst} reaches primitive heap assertions such as
$\code{arr}(p(i),l)$, it rewrites them using suitable lemmas (\eg,
turning $\code{arr}(p(i),l)$ into $\code{arr}(p(\phi(i)),l)$ via
\code{subst_array}), dispatching the residual goal by using Coq's
standard \code{reflexivity} tactic that instantiates the remaining
existential variable.





\section{\logic under the Spotlight: Empirical Evaluation}
\label{sec:eval}

We conducted an empirical evaluation of \logic to answer the following
research questions:


\begin{itemize}
\item \textbf{RQ1}: Is \logic expressive enough to reason about
  real-world computations on structured data?
\item \textbf{RQ2}: How \logic proofs are influenced by the involved
  formats and kinds of computations?
\item \textbf{RQ3}: How does \logic fare against state-of-the-art
  approaches not based on relational logic? 
\end{itemize}


\paragraph{Benchmarks} 

%
We have assembled our benchmark suite by adopting case studies from
different sources, including programs produced by the
TACO~\cite{KjolstadKCLA17} and \tname{Finch}~\cite{AhrensDKA23}
compilers, notable benchmarks from the Sparse Suite
collection~\cite{ChangwanRIS19}, and an example from a recent effort
on verifying sparse matrix/vector multiplication~\cite{LAproof}.
The conversion of the selected programs from their implementation
languages, C and Julia, to the language of \logic (\autoref{fig:prog})
has been done manually, but could be easily automated.
%
%

The resulting collection of 13~programs is summarised
in~\autoref{table:eval}.
The programs were selected to feature diverse range of data formats
and operations.
The considered formats can be broadly classified into three groups:
(1) \emph{primitive} sparse tensor formats, such by SV and Coordinate
List (COO)~\cite{popoola23},
(2) \emph{composite} sparse tensor data formats, CSR and uCSR, that
incorporate primitive formats as their parts,
and (3) a data format for non-sparse structured data, represented by
Run-Length Encoding (RL)~\cite{DonenfeldCA22}.
For each such format group, we considered the most common operations
with the respective data implemented in real-world kernels:
dot-product and alpha-blending for vector formats, and matrix/vector
multiplication for matrix formats.
All the verified programs listed in \autoref{table:eval} are
parametrised with arbitrary sizes of matrices/vectors.
%
%

\begin{table}[t]

  \caption{Statistics for the verified programs. For each program, we
    specify its operation in Einstein notation, the formats of the
    involved operands (D is for \emph{dense}), and the return type,
    followed by the size of the code and of the mechanised \logic
    proof, along with the proof/code ratio.
    We also indicate the characteristic rules of \logic the auxiliary
    specifications used by the proof, as well the time it takes Coq to
    check it.}
    
    \begin{minipage}{\textwidth}
    \centering
    \small{
    
    \resizebox{\columnwidth}{!}{%
    \begin{tabular}{c|c|c|c|cc|c|cccc|c|c}
    \toprule

    \multirow{2}{*}{{\#}} & \multirow{2}{*}{{Operation}} &
                                                           \multirow{2}{*}{{\makecell{Formats \\ (Operands)}}} & \multirow{2}{*}{{\makecell{Return \\ Type}}} & \multicolumn{2}{c|}{{Size (LOC)}} & \multirow{2}{*}{{Ratio}} & \multicolumn{4}{c|}{{Prominent rules}} & \multirow{2}{*}{{Uses}} & \multirow{2}{*}{{\makecell{Time \\ (sec)}}} \\[2pt]
    & & & & {Code} & {Proof} & & \rulename{For} & \rulename{While} & \rulename{Focus} & \rulename{Subst} & \\ \midrule
    
    1 & $\sum_{i}^{} x_{i}$ & {\textsf{SV} ($x$)}\textsuperscript{1,2} & {\texttt{int}} & 9 & 22 & 2.4 & \cmark &  & \cmark &  &  & 5.42 \\ 
    2 & $\sum_{i}^{} x_{i} y_{i}$ & {\textsf{SV} ($x$), \textsf{D} ($y$)}\textsuperscript{1,2} & {\texttt{int}} & 12 & 22 & 1.8 & \cmark &  & \cmark &  &  & 6.96 \\ 
    3 & $\sum_{i}^{} x_{i} y_{i}$ & {\textsf{SV} ($x$), \textsf{SV} ($y$)}\textsuperscript{1,2} & {\texttt{int}} & 31 & 237 & 7.6 &  & \cmark & \cmark &  &  & 20.73 \\ 
    4 & $\sum_{i, j}^{} A_{i, j}$ & {\textsf{COO} ($A$)} & {\texttt{int}} & 7 & 25 & 3.6 & \cmark &  & \cmark &  &  & 5.40 \\ 
    5 & $\sum_{i, j}^{} A_{i, j}$ & {\textsf{CSR} ($A$)}\textsuperscript{1,2} & {\texttt{int}} & 10 & 11 & 1.1 & \cmark &  &  & \cmark & \#1 & 7.91 \\ 
    6 & $\sum_{j}^{} A_{i, j} x_{j}$ & {\textsf{CSR} ($A$), \textsf{D} ($x$)}\textsuperscript{1,2} & {\texttt{int[]}} & 10 & 12 & 1.2 & \cmark &  &  & \cmark & \#2 & 8.99 \\ 
    7 & $\sum_{j}^{} A_{i, j} x_{j}$ & {\textsf{CSR} ($A$), \textsf{SV} ($x$)}\textsuperscript{1,2} & {\texttt{int[]}} & 10 & 16 & 1.6 & \cmark &  &  & \cmark & \#3 & 11.02 \\ 
    8 & $\sum_{i, j}^{} A_{i, j}$ & {\textsf{uCSR} ($A$)} & {\texttt{int}} & 10 & 29 & 2.9 & \cmark &  & \cmark & \cmark & \#1 & 12.62 \\ 
    9 & $\sum_{j}^{} A_{i, j} x_{j}$ & {\textsf{uCSR} ($A$), \textsf{D} ($x$)}\textsuperscript{3} & {\texttt{int[]}} & 11 & 33 & 3.0 & \cmark &  & \cmark & \cmark & \#2 & 18.69 \\ 
    10 & $\sum_{j}^{} A_{i, j} x_{j}$ & {\textsf{uCSR} ($A$), \textsf{SV} ($x$)}\textsuperscript{3} & {\texttt{int[]}} & 11 & 30 & 2.7 & \cmark &  & \cmark & \cmark & \#3 & 20.05 \\ 
    11 & $\sum_{i}^{} x_{i}$ & {\textsf{RL} ($x$)}\textsuperscript{2} & {\texttt{int}} & 16 & 18 & 1.1 & \cmark &  &  &  &  & 5.90 \\ 
    12 & $\sum_{i}^{} \alpha x_{i} + \beta y_{i}$ & {\textsf{RL} ($x$), \textsf{RL} ($y$)}\textsuperscript{2} & {\texttt{int}} & 64 & 143 & 2.2 &  & \cmark &  &  &  & 41.66 \\ 
    13 & $\sum_{j}^{} A_{i, j} x_{j}$ & {\textsf{CSR} ($A$), \textsf{D} ($x$)}\textsuperscript{4} & {\texttt{double[]}} & 19 & 36 & 1.9 & \cmark &  & \cmark & \cmark &  & 31.76 \\ 
    \bottomrule
    \multicolumn{13}{l}{}
    \\[-6pt]
    \multicolumn{13}{l} {
      \textsuperscript{1} From~\cite{KjolstadKCLA17}\qquad
      \textsuperscript{2} From~\cite{AhrensDKA23}\qquad
      \textsuperscript{3} From~\cite{ChangwanRIS19}\qquad
      \textsuperscript{4} From~\cite{LAproof}
      }
   
    \end{tabular}%
    }
    }
    \end{minipage}
    \label{table:eval}

\vspace{-15pt}
\end{table}



\subsection{RQ1: Expressivity of the Logic}

We answer RQ1 by providing quantitative evidence of \logic's
effectiveness and efficiency when verifying characteristic programs
manipulating with structured data listed in \autoref{table:eval}.

A quick glance at the table's fifth and sixth column should convey
that \logic's proofs are relatively compact, especially for an
extrinsic verifier, such as Coq, with the average proof/code ratio (in
LOC) being $2.5$.
One notable outlier in this aspect is a dot-product of two sparse
vectors~(\#3), whose proof we will discuss in \autoref{sec:trends}.
The last column of \autoref{table:eval} reports the proof checking
times (averaging at 15.2s), obtained via Coq 8.18 on a 3.2 GHz Apple
M1 MacBook Air with 16GB RAM.
The time it takes Coq to check \logic proofs is directly influenced by
the number of uses of tactics that implement domain-specific
automation, such as \code{xsubst} and \code{xlocal}
(\cf~\autoref{sec:auto}).


\subsection{RQ2: Verification Trends}
\label{sec:trends}

Besides quantitative characteristics of \logic, such as proof sizes
and proof checking time, we are also interested in \emph{verification
  trends}. For instance, what rules one should expect to use when
dealing with a particular format or a kind of structured data
computation?
%

To answer this question, we start by considering the programs \#1-\#4
involving primitive formats SV and COO for sparse vectors and
matrices, which are used as ``building blocks'' by other formats
The verification trends for these programs are summarised in the
seventh column of \autoref{table:eval}, featuring (1)~the
\rulename{Focus} rule, required to focus \emph{only on non-zero
  elements} of the input tensor and (2)~the \tname{For} or
\tname{While} rules, depending on the type of the loop in the code,
required to group remaining non-zero elements with correspondent loop
iterations.
Next, we focus on the case studies \#5-\#10 with composite formats,
CSR and uCSR, which are obtained as combinations of a dense
representation, SV (for exploiting sparcity), and COO (for featuring
random permutation of the components).
%
%
The key proof principle of \logic that shows up prominently in those
proofs is \tname{Subst}.
This should not come as a surprise: because of the composite nature of
the formats, one should expect to make use of the specifications that
involve their sub-formats (\eg, SV is a part of CSR).
Using \rulename{Subst} allows one to adjust intermediate subgoals
within the proof to match specifications of ``auxiliary'' operations
(most of which are represented by case studies \#1-\#4)
so they could be used in the proof in a compositional manner.
As the penultimate column of \autoref{table:eval} indicates, the case
studies \#5-\#10 that involve composite data formats, reuse a proof of
some case study \#1-\#3, involving only primitive formats.
Finally, we consider the case studies \#11-\#12, in which we have
verified operations on RL encoding that manipulate with compressed
arrays containing repeated data.
In those proofs, we made use only of \logic's loops rules, as the RL
format is primitive (so we don't need \tname{Subst}) and is not sparse
(so we don't need to focus on non-zero elements via \tname{Focus}).

We conclude this subsection with an observation about the only example
from our collection that has a relatively large proof/code ratio
(7.6): a dot-product of two SV vectors (\#3).
This figure can be explained by the fact that the textbook
implementation of SV/SV product is heavily optimised by
\emph{skipping} as many unnecessary computations (involving zeros) as
possible. Hence the main complexity in the proof is because of the
need to justify the \emph{absence} of code, rather than its
\emph{presence}.

\subsection{RQ3: Comparison with Non-Relational Proofs}
\label{sec:vst}

%

The landscape of tools for verified computations on structured data is
somewhat scarce, with the prior work mostly focusing on verified
compilers for sparse tensors~\cite{KovachPLDI23, DyerAB19,
  ArnoldHKBS10}. Those tools, which we discuss in detail in
\autoref{sec:related}, can only produce a subset of computations that
can be verified in LGTM, and they are not guaranteed to produce the
most performant format-specific code.
The various existing logics for $k$-safety either do not offer means
to work with array computations~\cite{SousaD16, dardinier2023hyper},
or are not even mechanised~\cite{DOsualdoFD22}, thus making it
impossible to conduct a meaningful comparison with them as tools.
Therefore, we only compare verification in \logic with formal
reasoning about structured data in a mechanises general-purpose
non-relational Separation Logic.

Recent work by~\citet{LAproof} presented a mechanised proof of a
sparse matrix/dense vector product function implemented in C, taking a
sparse matrix in the CSR format.
The C implementation is specified and verified using the Verified
Software Toolchain~(VST)~\cite{Appel:ESOP11}, a Separation Logic
framework embedded into Coq.
\citeauthor{LAproof}'s proofs follows a relatively conventional
\emph{two-layer paradigm} for verifying imperative
code~\cite{Appel22}: first a \emph{functional} model of sparse
matrices is defined, along with its operations (\eg, computing the
dot-product of a matrix row and a dense vector); next, the C function
is ascribed a specification in terms of a functional model, and the
verification amounts to proving that the imperative function
\emph{refines} the functional operation.

\begin{wrapfigure}[6]{r}{0.45\textwidth}
\vspace{-12pt}
\begin{minted}[numbers=left, xleftmargin=2.5ex, numbersep=6pt, escapeinside=@@, fontsize=\footnotesize]{python3}
def spmv(m_idx, m_ind, m_val, v):
  for i in range(0, length(m_idx)):
    for j in range(m_idx[i], m_idx[i+1]):
      ans[i] += m_val[j] * v[m_ind[j]]
\end{minted}
\setlength{\abovecaptionskip}{8pt}  
\caption{CSR matrix/dense vector product}
\label{fig:kellison}
\end{wrapfigure}

\autoref{fig:kellison} presents a slightly simplified CSR
sparse matrix/dense vector product implementation verified in
\citeauthor{LAproof}'s LAProof framework. The outer loop iterates
through the rows of a compressed CSR matrix \code{m} and the inner
loop sums up all non-zero~\code{m} elements in each row, multiplied by
the respective values of a dense vector~\code{v}.
The functional model of CSR is defined using a family of inductive
predicates.
First, the inductive predicate \code{crs_row_rep}
(\autoref{list:laproof-crs-row-rep}) relates the contents of a dense
vector and its sparse representation, capturing the shape of the SV
format.
Next, \code{crs_rep_aux} (\autoref{list:laproof-crs-rep-aux}) uses
\code{crs_row_rep} to express the relation between a dense matrix and
the corresponding CSR matrix.
Finally, an SL \emph{representation predicate} \code{crs_rep} (omitted
for brevity) relates a dense matrix to its sparse counterpart
(constrained by \code{crs_rep_aux}) stored as a C \code{struct} in the
memory.
The predicate \code{crs_rep} is used in the pre/postconditions of the
specification, and the postcondition states that the returned vector
is exactly the result of multiplying the input dense matrix and a
dense vector.
The LAproof verification proceeds as follows:

\begin{figure}[t]
\begin{tabular}{cc}

\begin{tabular}{c}
\begin{subfigure}{0.45\textwidth}
\begin{lstlisting}[language=Coq,xleftmargin=0pt,xrightmargin=0pt,basicstyle=\scriptsize\ttfamily,numbersep=5pt]
Inductive crs_row_rep {t} : forall cols (vals:list 
(ftype t)) col_ind (v:list (ftype t)), Prop :=
| crs_row_rep_nil: crs_row_rep 0%Z nil nil nil
| crs_row_rep_zero: forall cols vals col_ind v,
  crs_row_rep (cols-1) vals (map pred col_ind) v ->
  crs_row_rep cols vals col_ind (Zconst t 0 :: v)
| crs_row_rep_val: forall cols x vals col_ind v,
  crs_row_rep (cols-1) vals (map pred col_ind) v ->
  crs_row_rep cols (x::vals) (0::col_ind) (x::v).
\end{lstlisting}
\setlength{\abovecaptionskip}{0pt}
\caption{Inductive predicate for relating a vector \code{v} with length \code{cols} and its corresponding sparse vector represented by \code{col_ind} and \code{vals}. }
\label{list:laproof-crs-row-rep}
\end{subfigure}
\end{tabular}
&
\begin{tabular}{c}
\begin{subfigure}{0.45\textwidth}
\begin{lstlisting}[language=Coq,xleftmargin=0pt,xrightmargin=0pt,basicstyle=\scriptsize\ttfamily,numbersep=5pt]
Definition crs_rep_aux {t} (mval: matrix t) cols 
(vals: list (ftype t)) col_ind row_ptr: Prop :=
(* propositions about sortedness and lengths *) /\
forall j, 0 <= j < Zlength mval ->
crs_row_rep cols (sublist (Znth j row_ptr) 
  (Znth (j+1) row_ptr) vals)
(sublist (Znth j row_ptr) (Znth (j+1) row_ptr) 
  col_ind) (Znth j mval).
\end{lstlisting}
\setlength{\abovecaptionskip}{10pt}
\caption{Predicate for relating a dense matrix \code{mval} modelled as
  a list and a sparse matrix in CSR, represented by \code{row_ptr},
  \code{col_ind}, and \code{vals}. }
\label{list:laproof-crs-rep-aux}
\end{subfigure}
\end{tabular}

\end{tabular}

\setlength{\abovecaptionskip}{5pt}
\setlength{\belowcaptionskip}{-10pt}
\caption{Encoding of CSR format in VST by~\citet{LAproof};
  \texttt{ftype t} is a type for floating point numbers. }
\label{list:laproof-all}
\end{figure}

\begin{enumerate}
  \item Prove that the inner loop accumulates into \code{ans[i]} all values in the array \code{m_val} corresponding to the $i$\ith row, multiplied by respective values of v using the invariant
\vspace{-3pt}
\[
I(j) \triangleq \mathtt{ans}[i] = \sum^j_{k =
    \mathtt{m}_{\mathtt{idx}}[i]} \mathtt{m}_{\mathtt{val}}[k] \cdot
  v[\mathtt{m}_{\mathtt{ind}}[k]]
\]
\vspace{-3pt}
\item For any $i$ at the end of the loop, replace the sum at the right
  hand side of the equality in the invariant with the product of the
  corresponding rows of the logical ``dense'' matrix~\code{m} and the
  vector~\code{v}. This rewrite requires leveraging the properties of
  the representation predicate \code{csr_rep}, connecting the contents
  of \code{m_idx}, \code{m_ind}, and \code{m_val} with the elements of
  the logical matrix \code{m} (including zeros), via a set of
  predicate-specific lemmas.
\item Verify the outer loop using an invariant stating that at $i$\ith
  iteration, for each $k < i$,
  $\code{ans}[k] = \code{m}[k] \cdot \code{v}$, where \code{m[k]} is
  an $k$\ith row of the logical matrix \code{m}.
\end{enumerate}

\noindent
The most laborious part of this proof is its second step. It requires
a whole separate file with 350 LOC of Coq code to prove a number of
bespoke facts about \code{csr_rep}. Those facts are needed to
``align'' the representation predicate, which necessarily describes
the entire contents of the matrix \code{m}, with the logic of the
inner loop of the code above, which only operates on \code{m}'s
non-zero values.

In contrast, our approach does not require one to design such tailored
functional model for each new sparse format.
Instead, with \logic, one can use access functions (\eg,
\code{sv_get}) for interpreting the formats, relating their results to
that of the main function being specified. Those access functions can
be executed ``on demand'' in lockstep with the respective parts of the
operation's implementation and verified with a comparatively low proof
overhead.
This highlights a conceptual advantage of our approach: it liberates
the user from the burden associated with ($i$)~designing functional
format definitions and ($ii$)~proving their properties that are
typically required in refinement proofs.

To further demonstrate how our approach facilitates proofs about
sparse matrices, we conducted the verification task
from~\citeauthor{LAproof}'s work in \logic, which is marked as the
case study~\#13 in~\autoref{table:eval}.
In the interest of fair comparison, the program~\#13 differs from a
similar implementation~\#6 in the following aspects:
(a)~it works on the domain of IEEE 64-bit floating point numbers
instead of integers, and
(b) it does not call sub-functions (\eg,~\#2) to keep the full
resemblance to the C program by~\citeauthor{LAproof}, which does not
depend on sub-procedures that operate with sparse vectors.
To verify the inner loop of \code{smpv} in \autoref{fig:kellison} in
LGTM, we first replace the manipulations with the matrix contents via
\code{m_ind} and \code{m_val} by manipulations directly with the
contents of its logical row $\code{m}[i]$, so that the remaining goal
is conceptually reduced to a vector/vector product.
And then align each $i$\ith iteration of the inner loop with a
matching access function call for the $i$\ith non-zero element of the
matrix's row $\code{m}[i]$, and prove that the ``logical'' iterations
corresponding to zeros can be ignored. Verification of the outer loop
is similar to the one presented in \autoref{sec:overview}.

To summarise, we note that, instead of inventing a data representation
predicate and proving lemmas about~it, LGTM proof is mostly about
``aligning'' calls to access functions with the code being verified.
That is why, from~\autoref{table:eval}, our proof, even without
composition, is only 36~LOC.
In~\citeauthor{LAproof}'s work, the proofs about CSR-specific
properties alone take 210~LOC, and the refinement proof takes
additional 204~LOC, making our proof \emph{less than a tenth} of their
total length.



\section{Related Work}
\label{sec:related}

Our work connects several lines of research on (a)~program logics for
relational safety properties, (b)~compilers for structured data, and
(c)~verification of structured data manipulations.
%


\paragraph{Relational program logics.}
\citeauthor{Benton04}'s seminal work (\citeyear{Benton04}) introduced
Relational Hoare Logic~(RHL) to capture safety properties of
\emph{pairs of programs} and prove them using lockstep rules.
RHL's extension to Separation Logic~\cite{Yang07} has been used to
specify and verify complex relational properties of heap-manipulating
programs, such as equivalence of heap-based graph-marking algorithms.
\citet{BartheCK11} proposed a method to encode reasoning about
2-safety properties in unary Hoare logic by considering \emph{product
  programs} that simulate the execution steps of both their
constituents.
More recently, \citet{SousaD16} introduced Cartesian Hoare Logic (CHL)
that allows one to express and prove $k$-safety properties of
imperative programs for arbitrary but fixed~$k$.
CHL comes with the automated verifier \tname{Descartes} that proves
hypersafety specifications by reducing them to unary Hoare triples.
None of these approaches support reasoning about families of programs
indexed by elements of an arbitrary finite set that can change within
the proof.

A recent work by \citet{dardinier2023hyper} introduced Hyper Hoare
Logic (HHL)---a framework to reason about $k$-safety properties of
arbitrary (even infinite) arity~$k$.
Unlike \logic, HHL only allows for $k$-safety specifications
constraining multiple runs of \emph{the same} program.
Similarly to \logic, in HHL the arity of a judgement can change during
the proof, for instance in its \rulename{Cons} rule.
%
%
This feature is, however, not exploited by HHL, which does not provide
specialised rules for reasoning about loops, such as \logic's
\rulename{For} and \rulename{While}.
%
%
The closest to our work is Logic for Hyper-triple Composition (LHC)
by~\citet{DOsualdoFD22}. We provided a detailed discussion on the
improvements \logic makes over LHC in terms of expressivity in
\autoref{sec:kill-lhc}.





\paragraph{Compilers for structured data manipulations.}
Sparsity is a well-studied example of structured data, with
wide-ranging applications from machine learning to scientific
computing.
%
%
A popular strategy for sparse tensor compilation is to generate
low-level code directly from high-level abstractions
\cite{kjolstad_tensor_2019}. The TACO tensor compiler is emblematic of
the approach, defining sparse tensor formats through an iterator
interface for non-zero elements~\cite{ChouKA18, ChouKA20}, generating
an implementation of an operation for the sparse representations from
the same operation on dense arrays.
%
%
%
%
Later works generalised TACO's format abstraction to capture dynamic
data structures~\cite{ChouA22}, multiple simultaneous formats
\cite{ye_sparsetir_2022}, 
repetition~\cite{DonenfeldCA22}, or ragged-style
irregularity~\cite{FegadeCGM22}.
%
%
Many such approaches were unified and generalised by the \tname{Finch}
compiler~\cite{AhrensDKA23}, which goes beyond sparsity and allows one
to define arbitrary iteration strategies for producing efficient
looping code.






\paragraph{Verified computations with structured data.}
\citet{ArnoldHKBS10} proposed a high-level functional language LL for
expressing computations with sparse tensors that facilitates
automation of correctness proofs about matrix manipulations in
Isabelle/HOL.
The approach by~\citeauthor{ArnoldHKBS10} is, however, only limited to
sparse formats that can be encoded in LL.
Furthermore, it assumes correctness of the \emph{compilation} from the
LL encoding to the C representation~\cite{Arnold11}.
%
%
That is, the correctness guarantees provided by the approach only
apply to LL programs, but \emph{not} to their C counterparts.

%

\citet{DyerAB19} suggested to encode sparse formats and computation as
\tname{Alloy} predicates, phrasing verification of sparse tensor
computations as a search for counterexamples.
Such verification is unsound, in the sense that \tname{Alloy} tries to
find only small counterexamples up to a certain bound.
Moreover the way to encode stateful programs in \tname{Alloy} proposed
by \citeauthor{DyerAB19} is only capable to capture programs with
nested \code{for}-loops and could not express, \eg, \code{spmspv}
from~\autoref{sec:overview}.

\citet{LiuBCR22} developed a Coq library for certified optimisations
of tensor-manipulating program via verified rewrites.
Those rewrites do not exploit the structure of specific data formats,
and, hence, are not guaranteed to produce optimised implementations.
Finally \citet{KovachPLDI23} presented \tname{Etch}, a verified (in
Lean) compiler for sparse data computations.
\tname{Etch} is limited to operations on structures that are
expressible as mappings over indexed streams, so its optimisations can
be expressed as stream fusion, hierarchical iterations, and control
over interaction order~\cite{KiselyovBPS17}.
This restricts the range of operations \tname{Etch} can synthesise:
for instance, it cannot generate computations on data with unordered
layers, such as COO and uCSR formats.

Unlike the listed above efforts, \logic is not a compiler, but is a
verification framework: it does not synthesise computations, but
offers means to mechanically verify the results produced by highly
fine-tuned unverified synthesisers, such as TACO. This makes it a
suitable verification back-end for certifying sparse data compilers,
which are yet to be developed.
Furthermore, the applications of LGTM are not limited to just sparse
data: for example, our case studies \#11-12 from \autoref{table:eval}
are about the run-length encoding for arrays with repeated values.


\section{Conclusion and Future Work}
\label{sec:conclusion}

In this work we presented \logic: a program logic for hypersafety
properties, built around the idea of parametrised $k$-safety
specifications.
We argued that parametrisation enables intuitive safety proofs about
intricate computations over structured data (\eg, sparse tensors), and
substantiated this claim by mechanically verifying thirteen case
studies using LGTM embedding into Coq.
%

%

In our immediate future work, we are planning to automate common
reasoning patterns of \logic to facilitate push-the-button
construction of machine-assisted proofs for parametrised hypersafety
specifications.
%
%
As our long-term agenda, we are aiming to use \logic as a foundation
for building a \emph{certifying} compiler for structured data
manipulations similar to \tname{Finch}~\cite{AhrensDKA23} by adopting
the ideas of proof-producing program synthesis~\cite{WatanabeGPPS21}.


%
%

\begin{acks}

We thank Kiran Gopinathan, Yunjeong Lee, Peter M\"{u}ller, and
George P\^{i}rlea for their feedback on drafts of this paper.
We also thank the anonymous PLDI'24 PC and AEC reviewers for their
constructive and insightful comments.
This work was partially supported by a Singapore Ministry of Education
(MoE) Tier 3 grant ``Automated Program Repair'' MOE-MOET32021-0001.
\end{acks}

\section*{Data Availability}

The software artefact accompanying this paper is available
online~\cite{lgtm}.
The artefact contains the source code and build scripts for the Coq
mechanisation of \logic, as well as the collection of case studies
that can be used to reproduce the results described in
\autoref{sec:eval}.

\bibliography{references}


\begin{thebibliography}{36}


\ifx \showCODEN    \undefined \def \showCODEN     #1{\unskip}     \fi
\ifx \showDOI      \undefined \def \showDOI       #1{#1}\fi
\ifx \showISBNx    \undefined \def \showISBNx     #1{\unskip}     \fi
\ifx \showISBNxiii \undefined \def \showISBNxiii  #1{\unskip}     \fi
\ifx \showISSN     \undefined \def \showISSN      #1{\unskip}     \fi
\ifx \showLCCN     \undefined \def \showLCCN      #1{\unskip}     \fi
\ifx \shownote     \undefined \def \shownote      #1{#1}          \fi
\ifx \showarticletitle \undefined \def \showarticletitle #1{#1}   \fi
\ifx \showURL      \undefined \def \showURL       {\relax}        \fi
\providecommand\bibfield[2]{#2}
\providecommand\bibinfo[2]{#2}
\providecommand\natexlab[1]{#1}
\providecommand\showeprint[2][]{arXiv:#2}

\bibitem[Ahrens et~al\mbox{.}(2023)]%
        {AhrensDKA23}
\bibfield{author}{\bibinfo{person}{Willow Ahrens}, \bibinfo{person}{Daniel
  Donenfeld}, \bibinfo{person}{Fredrik Kjolstad}, {and}
  \bibinfo{person}{Saman~P. Amarasinghe}.} \bibinfo{year}{2023}\natexlab{}.
\newblock \showarticletitle{Looplets: {A} Language for Structured Coiteration}.
  In \bibinfo{booktitle}{\emph{CGO}}. \bibinfo{publisher}{{ACM}},
  \bibinfo{pages}{41--54}.
\newblock
\urldef\tempurl%
\url{https://doi.org/10.1145/3579990.3580020}
\showDOI{\tempurl}


\bibitem[Appel(2011)]%
        {Appel:ESOP11}
\bibfield{author}{\bibinfo{person}{Andrew~W. Appel}.}
  \bibinfo{year}{2011}\natexlab{}.
\newblock \showarticletitle{{Verified Software Toolchain - (Invited Talk)}}. In
  \bibinfo{booktitle}{\emph{ESOP}} \emph{(\bibinfo{series}{LNCS},
  Vol.~\bibinfo{volume}{6602})}. \bibinfo{publisher}{Springer},
  \bibinfo{pages}{1--17}.
\newblock
\urldef\tempurl%
\url{https://doi.org/10.1007/978-3-642-19718-5\_1}
\showDOI{\tempurl}


\bibitem[Appel(2022)]%
        {Appel22}
\bibfield{author}{\bibinfo{person}{Andrew~W. Appel}.}
  \bibinfo{year}{2022}\natexlab{}.
\newblock \showarticletitle{Coq’s vibrant ecosystem for verification
  engineering (invited talk)}. In \bibinfo{booktitle}{\emph{CPP}}.
  \bibinfo{publisher}{{ACM}}, \bibinfo{pages}{2--11}.
\newblock
\urldef\tempurl%
\url{https://doi.org/10.1145/3497775.3503951}
\showDOI{\tempurl}


\bibitem[Arnold(2011)]%
        {Arnold11}
\bibfield{author}{\bibinfo{person}{Gilad Arnold}.}
  \bibinfo{year}{2011}\natexlab{}.
\newblock \emph{\bibinfo{title}{Data-Parallel Language for Correct and
  Efficient Sparse Matrix Codes}}.
\newblock \bibinfo{thesistype}{Ph.\,D. Dissertation}.
  \bibinfo{school}{University of California, Berkeley, {USA}}.
\newblock
\urldef\tempurl%
\url{http://www.escholarship.org/uc/item/2pw6165p}
\showURL{%
\tempurl}


\bibitem[Arnold et~al\mbox{.}(2010)]%
        {ArnoldHKBS10}
\bibfield{author}{\bibinfo{person}{Gilad Arnold}, \bibinfo{person}{Johannes
  H{\"{o}}lzl}, \bibinfo{person}{Ali~Sinan K{\"{o}}ksal},
  \bibinfo{person}{Rastislav Bod{\'{\i}}k}, {and} \bibinfo{person}{Mooly
  Sagiv}.} \bibinfo{year}{2010}\natexlab{}.
\newblock \showarticletitle{Specifying and verifying sparse matrix codes}. In
  \bibinfo{booktitle}{\emph{ICFP}}. \bibinfo{publisher}{{ACM}},
  \bibinfo{pages}{249--260}.
\newblock
\urldef\tempurl%
\url{https://doi.org/10.1145/1863543.1863581}
\showDOI{\tempurl}


\bibitem[Barthe et~al\mbox{.}(2011)]%
        {BartheCK11}
\bibfield{author}{\bibinfo{person}{Gilles Barthe}, \bibinfo{person}{Juan~Manuel
  Crespo}, {and} \bibinfo{person}{C{\'{e}}sar Kunz}.}
  \bibinfo{year}{2011}\natexlab{}.
\newblock \showarticletitle{{Relational Verification Using Product Programs}}.
  In \bibinfo{booktitle}{\emph{{FM}}} \emph{(\bibinfo{series}{LNCS},
  Vol.~\bibinfo{volume}{6664})}. \bibinfo{publisher}{Springer},
  \bibinfo{pages}{200--214}.
\newblock
\urldef\tempurl%
\url{https://doi.org/10.1007/978-3-642-21437-0\_17}
\showDOI{\tempurl}


\bibitem[Barthe et~al\mbox{.}(2012)]%
        {BartheKOB12}
\bibfield{author}{\bibinfo{person}{Gilles Barthe}, \bibinfo{person}{Boris
  K{\"{o}}pf}, \bibinfo{person}{Federico Olmedo}, {and}
  \bibinfo{person}{Santiago~Zanella B{\'{e}}guelin}.}
  \bibinfo{year}{2012}\natexlab{}.
\newblock \showarticletitle{Probabilistic relational reasoning for differential
  privacy}. In \bibinfo{booktitle}{\emph{POPL}}. \bibinfo{publisher}{{ACM}},
  \bibinfo{pages}{97--110}.
\newblock
\urldef\tempurl%
\url{https://doi.org/10.1145/2103656.2103670}
\showDOI{\tempurl}


\bibitem[Benton(2004)]%
        {Benton04}
\bibfield{author}{\bibinfo{person}{Nick Benton}.}
  \bibinfo{year}{2004}\natexlab{}.
\newblock \showarticletitle{{Simple Relational Correctness Proofs for Static
  Analyses and Program Transformations}}. In \bibinfo{booktitle}{\emph{POPL}}.
  \bibinfo{publisher}{{ACM}}, \bibinfo{pages}{14--25}.
\newblock
\urldef\tempurl%
\url{https://doi.org/10.1145/964001.964003}
\showDOI{\tempurl}


\bibitem[Carbin et~al\mbox{.}(2012)]%
        {CarbinKMR12}
\bibfield{author}{\bibinfo{person}{Michael Carbin}, \bibinfo{person}{Deokhwan
  Kim}, \bibinfo{person}{Sasa Misailovic}, {and} \bibinfo{person}{Martin~C.
  Rinard}.} \bibinfo{year}{2012}\natexlab{}.
\newblock \showarticletitle{Proving acceptability properties of relaxed
  nondeterministic approximate programs}. In \bibinfo{booktitle}{\emph{PLDI}}.
  \bibinfo{publisher}{{ACM}}, \bibinfo{pages}{169--180}.
\newblock
\urldef\tempurl%
\url{https://doi.org/10.1145/2254064.2254086}
\showDOI{\tempurl}


\bibitem[Chargu{\'{e}}raud(2011)]%
        {Chargueraud11}
\bibfield{author}{\bibinfo{person}{Arthur Chargu{\'{e}}raud}.}
  \bibinfo{year}{2011}\natexlab{}.
\newblock \showarticletitle{{Characteristic Formulae for the Verification of
  Imperative Programs}}. In \bibinfo{booktitle}{\emph{ICFP}}.
  \bibinfo{publisher}{{ACM}}, \bibinfo{pages}{418--430}.
\newblock
\urldef\tempurl%
\url{https://doi.org/10.1145/2034773.2034828}
\showDOI{\tempurl}


\bibitem[Chargu{\'{e}}raud(2020)]%
        {Chargueraud20}
\bibfield{author}{\bibinfo{person}{Arthur Chargu{\'{e}}raud}.}
  \bibinfo{year}{2020}\natexlab{}.
\newblock \showarticletitle{{Separation Logic for Sequential Programs
  (Functional Pearl)}}.
\newblock \bibinfo{journal}{\emph{Proc. {ACM} Program. Lang.}}
  \bibinfo{volume}{4}, \bibinfo{number}{{ICFP}} (\bibinfo{year}{2020}),
  \bibinfo{pages}{116:1--116:34}.
\newblock
\urldef\tempurl%
\url{https://doi.org/10.1145/3408998}
\showDOI{\tempurl}


\bibitem[Chou and Amarasinghe(2022)]%
        {ChouA22}
\bibfield{author}{\bibinfo{person}{Stephen Chou} {and}
  \bibinfo{person}{Saman~P. Amarasinghe}.} \bibinfo{year}{2022}\natexlab{}.
\newblock \showarticletitle{Compilation of dynamic sparse tensor algebra}.
\newblock \bibinfo{journal}{\emph{Proc. {ACM} Program. Lang.}}
  \bibinfo{volume}{6}, \bibinfo{number}{{OOPSLA2}} (\bibinfo{year}{2022}),
  \bibinfo{pages}{1408--1437}.
\newblock
\urldef\tempurl%
\url{https://doi.org/10.1145/3563338}
\showDOI{\tempurl}


\bibitem[Chou et~al\mbox{.}(2018)]%
        {ChouKA18}
\bibfield{author}{\bibinfo{person}{Stephen Chou}, \bibinfo{person}{Fredrik
  Kjolstad}, {and} \bibinfo{person}{Saman~P. Amarasinghe}.}
  \bibinfo{year}{2018}\natexlab{}.
\newblock \showarticletitle{Format abstraction for sparse tensor algebra
  compilers}.
\newblock \bibinfo{journal}{\emph{Proc. {ACM} Program. Lang.}}
  \bibinfo{volume}{2}, \bibinfo{number}{{OOPSLA}} (\bibinfo{year}{2018}),
  \bibinfo{pages}{123:1--123:30}.
\newblock
\urldef\tempurl%
\url{https://doi.org/10.1145/3276493}
\showDOI{\tempurl}


\bibitem[Chou et~al\mbox{.}(2020)]%
        {ChouKA20}
\bibfield{author}{\bibinfo{person}{Stephen Chou}, \bibinfo{person}{Fredrik
  Kjolstad}, {and} \bibinfo{person}{Saman~P. Amarasinghe}.}
  \bibinfo{year}{2020}\natexlab{}.
\newblock \showarticletitle{Automatic generation of efficient sparse tensor
  format conversion routines}. In \bibinfo{booktitle}{\emph{PLDI}}.
  \bibinfo{publisher}{{ACM}}, \bibinfo{pages}{823--838}.
\newblock
\urldef\tempurl%
\url{https://doi.org/10.1145/3385412.3385963}
\showDOI{\tempurl}


\bibitem[Dardinier and M{\"{u}}ller(2023)]%
        {dardinier2023hyper}
\bibfield{author}{\bibinfo{person}{Thibault Dardinier} {and}
  \bibinfo{person}{Peter M{\"{u}}ller}.} \bibinfo{year}{2023}\natexlab{}.
\newblock \showarticletitle{{Hyper Hoare Logic: (Dis-)Proving Program
  Hyperproperties (extended version)}}.
\newblock \bibinfo{journal}{\emph{CoRR}}  \bibinfo{volume}{abs/2301.10037}
  (\bibinfo{year}{2023}).
\newblock
\urldef\tempurl%
\url{https://doi.org/10.48550/arXiv.2301.10037}
\showDOI{\tempurl}


\bibitem[Donenfeld et~al\mbox{.}(2022)]%
        {DonenfeldCA22}
\bibfield{author}{\bibinfo{person}{Daniel Donenfeld}, \bibinfo{person}{Stephen
  Chou}, {and} \bibinfo{person}{Saman~P. Amarasinghe}.}
  \bibinfo{year}{2022}\natexlab{}.
\newblock \showarticletitle{{Unified Compilation for Lossless Compression and
  Sparse Computing}}. In \bibinfo{booktitle}{\emph{CGO}}.
  \bibinfo{publisher}{{IEEE}}, \bibinfo{pages}{205--216}.
\newblock
\urldef\tempurl%
\url{https://doi.org/10.1109/CGO53902.2022.9741282}
\showDOI{\tempurl}


\bibitem[D'Osualdo et~al\mbox{.}(2022)]%
        {DOsualdoFD22}
\bibfield{author}{\bibinfo{person}{Emanuele D'Osualdo}, \bibinfo{person}{Azadeh
  Farzan}, {and} \bibinfo{person}{Derek Dreyer}.}
  \bibinfo{year}{2022}\natexlab{}.
\newblock \showarticletitle{Proving hypersafety compositionally}.
\newblock \bibinfo{journal}{\emph{Proc. {ACM} Program. Lang.}}
  \bibinfo{volume}{6}, \bibinfo{number}{{OOPSLA2}} (\bibinfo{year}{2022}),
  \bibinfo{pages}{289--314}.
\newblock
\urldef\tempurl%
\url{https://doi.org/10.1145/3563298}
\showDOI{\tempurl}


\bibitem[Dyer et~al\mbox{.}(2019)]%
        {DyerAB19}
\bibfield{author}{\bibinfo{person}{Tristan Dyer}, \bibinfo{person}{Alper
  Altuntas}, {and} \bibinfo{person}{John W.~Baugh Jr.}}
  \bibinfo{year}{2019}\natexlab{}.
\newblock \showarticletitle{Bounded Verification of Sparse Matrix
  Computations}. In \bibinfo{booktitle}{\emph{2019 {IEEE/ACM} 3rd International
  Workshop on Software Correctness for {HPC} Applications (Correctness)}}.
  \bibinfo{publisher}{{IEEE}}, \bibinfo{pages}{36--43}.
\newblock
\urldef\tempurl%
\url{https://doi.org/10.1109/Correctness49594.2019.00010}
\showDOI{\tempurl}


\bibitem[Fegade et~al\mbox{.}(2022)]%
        {FegadeCGM22}
\bibfield{author}{\bibinfo{person}{Pratik Fegade}, \bibinfo{person}{Tianqi
  Chen}, \bibinfo{person}{Phillip~B. Gibbons}, {and} \bibinfo{person}{Todd~C.
  Mowry}.} \bibinfo{year}{2022}\natexlab{}.
\newblock \showarticletitle{{The CoRa Tensor Compiler: Compilation for Ragged
  Tensors with Minimal Padding}}. In \bibinfo{booktitle}{\emph{MLSys}}.
  \bibinfo{publisher}{mlsys.org}.
\newblock
\urldef\tempurl%
\url{https://proceedings.mlsys.org/paper/2022/hash/d3d9446802a44259755d38e6d163e820-Abstract.html}
\showURL{%
\tempurl}


\bibitem[Gladshtein et~al\mbox{.}(2024)]%
        {lgtm}
\bibfield{author}{\bibinfo{person}{Vladimir Gladshtein},
  \bibinfo{person}{Qiyuan Zhao}, \bibinfo{person}{Willow Ahrens},
  \bibinfo{person}{Saman Amarasinghe}, {and} \bibinfo{person}{Ilya Sergey}.}
  \bibinfo{year}{2024}\natexlab{}.
\newblock \bibinfo{booktitle}{\emph{LGTM: the Logic for Graceful Tensor
  Manipulation}}.
\newblock
\urldef\tempurl%
\url{https://doi.org/10.5281/zenodo.10802914}
\showDOI{\tempurl}


\bibitem[Hong et~al\mbox{.}(2019)]%
        {ChangwanRIS19}
\bibfield{author}{\bibinfo{person}{Changwan Hong}, \bibinfo{person}{Aravind
  Sukumaran~Rajam}, \bibinfo{person}{Israt Nisa}, \bibinfo{person}{Kunal
  Singh}, {and} \bibinfo{person}{Ponnuswamy Sadayappan}.}
  \bibinfo{year}{2019}\natexlab{}.
\newblock \showarticletitle{Adaptive sparse tiling for sparse matrix
  multiplication}. In \bibinfo{booktitle}{\emph{PPoPP}}.
  \bibinfo{publisher}{{ACM}}, \bibinfo{pages}{300--314}.
\newblock
\urldef\tempurl%
\url{https://doi.org/10.1145/3293883.3295712}
\showDOI{\tempurl}


\bibitem[Kellison et~al\mbox{.}(2023)]%
        {LAproof}
\bibfield{author}{\bibinfo{person}{Ariel~E. Kellison},
  \bibinfo{person}{Andrew~W. Appel}, \bibinfo{person}{Mohit Tekriwal}, {and}
  \bibinfo{person}{David~S. Bindel}.} \bibinfo{year}{2023}\natexlab{}.
\newblock \showarticletitle{{LAProof}: A Library of Formal Proofs of Accuracy
  and Correctness for Linear Algebra Programs}. In
  \bibinfo{booktitle}{\emph{Proceedings of the 30th IEEE International
  Symposium on Computer Arithhmetic (ARITH)}}.
\newblock


\bibitem[Kiselyov et~al\mbox{.}(2017)]%
        {KiselyovBPS17}
\bibfield{author}{\bibinfo{person}{Oleg Kiselyov}, \bibinfo{person}{Aggelos
  Biboudis}, \bibinfo{person}{Nick Palladinos}, {and} \bibinfo{person}{Yannis
  Smaragdakis}.} \bibinfo{year}{2017}\natexlab{}.
\newblock \showarticletitle{Stream fusion, to completeness}. In
  \bibinfo{booktitle}{\emph{POPL}}. \bibinfo{publisher}{{ACM}},
  \bibinfo{pages}{285--299}.
\newblock
\urldef\tempurl%
\url{https://doi.org/10.1145/3009837.3009880}
\showDOI{\tempurl}


\bibitem[Kjolstad et~al\mbox{.}(2019)]%
        {kjolstad_tensor_2019}
\bibfield{author}{\bibinfo{person}{Fredrik Kjolstad}, \bibinfo{person}{Willow
  Ahrens}, \bibinfo{person}{Shoaib Kamil}, {and} \bibinfo{person}{Saman
  Amarasinghe}.} \bibinfo{year}{2019}\natexlab{}.
\newblock \showarticletitle{Tensor {Algebra} {Compilation} with {Workspaces}}.
  In \bibinfo{booktitle}{\emph{CGO}}. \bibinfo{pages}{180--192}.
\newblock
\urldef\tempurl%
\url{https://doi.org/10.1109/CGO.2019.8661185}
\showDOI{\tempurl}


\bibitem[Kjolstad et~al\mbox{.}(2017)]%
        {KjolstadKCLA17}
\bibfield{author}{\bibinfo{person}{Fredrik Kjolstad}, \bibinfo{person}{Shoaib
  Kamil}, \bibinfo{person}{Stephen Chou}, \bibinfo{person}{David Lugato}, {and}
  \bibinfo{person}{Saman~P. Amarasinghe}.} \bibinfo{year}{2017}\natexlab{}.
\newblock \showarticletitle{The tensor algebra compiler}.
\newblock \bibinfo{journal}{\emph{Proc. {ACM} Program. Lang.}}
  \bibinfo{volume}{1}, \bibinfo{number}{{OOPSLA}} (\bibinfo{year}{2017}),
  \bibinfo{pages}{77:1--77:29}.
\newblock
\urldef\tempurl%
\url{https://doi.org/10.1145/3133901}
\showDOI{\tempurl}


\bibitem[Kovach et~al\mbox{.}(2023)]%
        {KovachPLDI23}
\bibfield{author}{\bibinfo{person}{Scott Kovach}, \bibinfo{person}{Praneeth
  Kolichala}, \bibinfo{person}{Tiancheng Gu}, {and} \bibinfo{person}{Fredrik
  Kjolstad}.} \bibinfo{year}{2023}\natexlab{}.
\newblock \showarticletitle{Indexed Streams: A Formal Intermediate
  Representation for Fused Contraction Programs}.
\newblock \bibinfo{journal}{\emph{Proc. ACM Program. Lang.}}
  \bibinfo{volume}{7}, \bibinfo{number}{PLDI}, Article \bibinfo{articleno}{154}
  (\bibinfo{year}{2023}).
\newblock
\urldef\tempurl%
\url{https://doi.org/10.1145/3591268}
\showDOI{\tempurl}


\bibitem[Liu et~al\mbox{.}(2022)]%
        {LiuBCR22}
\bibfield{author}{\bibinfo{person}{Amanda Liu}, \bibinfo{person}{Gilbert~Louis
  Bernstein}, \bibinfo{person}{Adam Chlipala}, {and} \bibinfo{person}{Jonathan
  Ragan{-}Kelley}.} \bibinfo{year}{2022}\natexlab{}.
\newblock \showarticletitle{Verified tensor-program optimization via high-level
  scheduling rewrites}.
\newblock \bibinfo{journal}{\emph{Proc. {ACM} Program. Lang.}}
  \bibinfo{volume}{6}, \bibinfo{number}{{POPL}} (\bibinfo{year}{2022}),
  \bibinfo{pages}{1--28}.
\newblock
\urldef\tempurl%
\url{https://doi.org/10.1145/3498717}
\showDOI{\tempurl}


\bibitem[Mahboubi and Tassi(2022)]%
        {Maboubi-Tassi:MathComp}
\bibfield{author}{\bibinfo{person}{Assia Mahboubi} {and}
  \bibinfo{person}{Enrico Tassi}.} \bibinfo{year}{2022}\natexlab{}.
\newblock \bibinfo{booktitle}{\emph{Mathematical Components}}.
\newblock
\urldef\tempurl%
\url{https://doi.org/10.5281/zenodo.7118596}
\showDOI{\tempurl}


\bibitem[Nanevski et~al\mbox{.}(2011)]%
        {NanevskiBG11}
\bibfield{author}{\bibinfo{person}{Aleksandar Nanevski},
  \bibinfo{person}{Anindya Banerjee}, {and} \bibinfo{person}{Deepak Garg}.}
  \bibinfo{year}{2011}\natexlab{}.
\newblock \showarticletitle{{Verification of Information Flow and Access
  Control Policies with Dependent Types}}. In \bibinfo{booktitle}{\emph{32nd
  {IEEE} Symposium on Security and Privacy}}. \bibinfo{publisher}{{IEEE}
  Computer Society}, \bibinfo{pages}{165--179}.
\newblock
\urldef\tempurl%
\url{https://doi.org/10.1109/SP.2011.12}
\showDOI{\tempurl}


\bibitem[Popoola et~al\mbox{.}(2023)]%
        {popoola23}
\bibfield{author}{\bibinfo{person}{Tobi Popoola}, \bibinfo{person}{Tuowen
  Zhao}, \bibinfo{person}{Aaron St.~George}, \bibinfo{person}{Kalyan Bhetwal},
  \bibinfo{person}{Michelle~Mills Strout}, \bibinfo{person}{Mary Hall}, {and}
  \bibinfo{person}{Catherine Olschanowsky}.} \bibinfo{year}{2023}\natexlab{}.
\newblock \showarticletitle{Code Synthesis for Sparse Tensor Format Conversion
  and Optimization}. In \bibinfo{booktitle}{\emph{CGO}}.
  \bibinfo{publisher}{ACM}, \bibinfo{pages}{28–40}.
\newblock
\urldef\tempurl%
\url{https://doi.org/10.1145/3579990.3580021}
\showDOI{\tempurl}


\bibitem[Reynolds(2002)]%
        {Reynolds:LICS02}
\bibfield{author}{\bibinfo{person}{John~C. Reynolds}.}
  \bibinfo{year}{2002}\natexlab{}.
\newblock \showarticletitle{Separation Logic: {A} Logic for Shared Mutable Data
  Structures}. In \bibinfo{booktitle}{\emph{LICS}}. \bibinfo{publisher}{{IEEE}
  Computer Society}, \bibinfo{pages}{55--74}.
\newblock
\urldef\tempurl%
\url{https://doi.org/10.1109/LICS.2002.1029817}
\showDOI{\tempurl}


\bibitem[Sakka et~al\mbox{.}(2017)]%
        {SakkaS017}
\bibfield{author}{\bibinfo{person}{Laith Sakka}, \bibinfo{person}{Kirshanthan
  Sundararajah}, {and} \bibinfo{person}{Milind Kulkarni}.}
  \bibinfo{year}{2017}\natexlab{}.
\newblock \showarticletitle{{TreeFuser: a framework for analyzing and fusing
  general recursive tree traversals}}.
\newblock \bibinfo{journal}{\emph{Proc. {ACM} Program. Lang.}}
  \bibinfo{volume}{1}, \bibinfo{number}{{OOPSLA}} (\bibinfo{year}{2017}),
  \bibinfo{pages}{76:1--76:30}.
\newblock
\urldef\tempurl%
\url{https://doi.org/10.1145/3133900}
\showDOI{\tempurl}


\bibitem[Sousa and Dillig(2016)]%
        {SousaD16}
\bibfield{author}{\bibinfo{person}{Marcelo Sousa} {and} \bibinfo{person}{Isil
  Dillig}.} \bibinfo{year}{2016}\natexlab{}.
\newblock \showarticletitle{{Cartesian Hoare Logic for Verifying k-Safety
  Properties}}. In \bibinfo{booktitle}{\emph{PLDI}}.
  \bibinfo{publisher}{{ACM}}, \bibinfo{pages}{57--69}.
\newblock
\urldef\tempurl%
\url{https://doi.org/10.1145/2908080.2908092}
\showDOI{\tempurl}


\bibitem[Watanabe et~al\mbox{.}(2021)]%
        {WatanabeGPPS21}
\bibfield{author}{\bibinfo{person}{Yasunari Watanabe}, \bibinfo{person}{Kiran
  Gopinathan}, \bibinfo{person}{George P{\^{\i}}rlea}, \bibinfo{person}{Nadia
  Polikarpova}, {and} \bibinfo{person}{Ilya Sergey}.}
  \bibinfo{year}{2021}\natexlab{}.
\newblock \showarticletitle{{Certifying the Synthesis of Heap-Manipulating
  Programs}}.
\newblock \bibinfo{journal}{\emph{Proc. {ACM} Program. Lang.}}
  \bibinfo{volume}{5}, \bibinfo{number}{{ICFP}} (\bibinfo{year}{2021}),
  \bibinfo{pages}{1--29}.
\newblock
\urldef\tempurl%
\url{https://doi.org/10.1145/3473589}
\showDOI{\tempurl}


\bibitem[Yang(2007)]%
        {Yang07}
\bibfield{author}{\bibinfo{person}{Hongseok Yang}.}
  \bibinfo{year}{2007}\natexlab{}.
\newblock \showarticletitle{Relational separation logic}.
\newblock \bibinfo{journal}{\emph{Theor. Comput. Sci.}} \bibinfo{volume}{375},
  \bibinfo{number}{1-3} (\bibinfo{year}{2007}), \bibinfo{pages}{308--334}.
\newblock
\urldef\tempurl%
\url{https://doi.org/10.1016/j.tcs.2006.12.036}
\showDOI{\tempurl}


\bibitem[Ye et~al\mbox{.}(2023)]%
        {ye_sparsetir_2022}
\bibfield{author}{\bibinfo{person}{Zihao Ye}, \bibinfo{person}{Ruihang Lai},
  \bibinfo{person}{Junru Shao}, \bibinfo{person}{Tianqi Chen}, {and}
  \bibinfo{person}{Luis Ceze}.} \bibinfo{year}{2023}\natexlab{}.
\newblock \showarticletitle{{SparseTIR: Composable Abstractions for Sparse
  Compilation in Deep Learning}}. In \bibinfo{booktitle}{\emph{ASPLOS}}.
  \bibinfo{publisher}{{ACM}}, \bibinfo{pages}{660--678}.
\newblock
\urldef\tempurl%
\url{https://doi.org/10.1145/3582016.3582047}
\showDOI{\tempurl}


\end{thebibliography}

\setlength\floatsep{1.25\baselineskip plus 3pt minus 2pt}
\setlength\textfloatsep{1.25\baselineskip plus 3pt minus 2pt}
\setlength\intextsep{1.25\baselineskip plus 3pt minus 2 pt}



\appendix

\section{\logic Rules}
\label{sec:appendix}


\EnableBpAbbreviations

\paragraph{Structural Rules}
\label{sec:structural}

\begin{figure}[h]
\setlength{\belowcaptionskip}{-5pt}
{\small{
  \begin{subfigure}{\textwidth}
    \begin{prooftree}
      \AXC{\(S_1\cap S_2 = \varnothing \)}
      \RightLabel{\tname{WpNest}}
      \UIC{\(\wp{S_1:\pp_1,\ S_2:\pp_2}{Q} \bientails \wpf{S_1:\ \pp_1}{\many{v}}{{\color{black}\wpf{S_2:\ \pp_2}{\many{u}}{Q(\many{vu})}}} \)}
    \end{prooftree}
  \end{subfigure}
  \begin{subfigure}{\textwidth}
    \vspace{2mm}
    \begin{prooftree}
      \AXC{\(\local{\{P_i, Q_i\}}{i}\)}
      \AXC{\(\forall i, \hsll{\pure{i \in S} \str P_i}{i:\ \pp(i)}{x}{Q_i(x)}\)}
      \RightLabel{\tname{Product}}
      \BIC{\(\hslc{\bigstr_{i\in S} P_i}{S:\ \pp}{\many{x}}{\bigstr_{i\in S} Q_i(\many{x}(i))}\)}
    \end{prooftree}
  \end{subfigure}
  \begin{subfigure}{0.50\textwidth}
    \vspace{2mm}
    \begin{prooftree}
      \AXC{\(H\vdash H' \quad Q'\vdash Q \quad \hslcc{H'}{S:\pp}{Q'}\)}
      \RightLabel{\tname{Conseq}}
      \UIC{\(\hslcc{H}{S:\pp}{Q}\)}
    \end{prooftree}
  \end{subfigure}
  \begin{subfigure}{0.33\textwidth}
    \vspace{2mm}
    \begin{prooftree}
      \AXC{\(\hslcc{H}{S:\pp}{Q}\)}
      \RightLabel{\tname{Frame}}
      \UIC{\(\hslcc{H'\!\str\! H}{S:\pp}{Q\!\str\! H'}\)}
    \end{prooftree}
  \end{subfigure}
  \begin{subfigure}{\textwidth}
    \vspace{2mm}
    \begin{prooftree}
      \AXC{\(\hsll{\! P\! }{S_1:\ \pp_1}{\many{x}}{H(\many{x}) \!\! }\)}      
      \AXC{\(\forall \many{x},\ \hsll{\! H(\many{x}) \! }{S_2:\ \pp_2}{\many{y}}{Q(\many{x}\many{y}) \!\! }\)}
      \RightLabel{\tname{SeqU}}
      \BIC{\(\hsll{\! P\! }{S_1:\ \pp_1,\ S_2:\ \pp_2}{\many{z}}{Q(\many{z}) \!\ }\)}
    \end{prooftree}  
  \end{subfigure}

  \begin{minipage}{0.48\textwidth}
    {\small{
  \begin{mathpar}
  \!\!\!\!\!\!\!\!\!\!\!\!
  \inferrule[\tname{SeqU1}]
  {
  \begin{array}{c}
    \hsll{\! P\! }{\iota:\ p_1}{{x}}{H \!\! }
    \arcr[2pt]
    \hsllazy{\! H \ \ }{\iota:\ p'_1, \ S:\ \pp_2\ }{{x}, \many{z}}{Q(\many{z}) \!\! }    
  \end{array}
  }
  {
  \hsllazy{\! P\ \ }{\iota : p_1; p'_1,~~S : \pp_2\ }{{x},
    \many{y}}{Q(\many{x}\many{y}) \!\! }
  }
  \end{mathpar}
  }}
  \end{minipage}
  \!\!\!\!
  \begin{minipage}{0.48\textwidth}
    {\small{  
  \inferrule[\tname{SeqU2}]
  {
  \begin{array}{c}
  \hsllazy{\!\! P \ }{\iota:\ p_1,\ S:\ \pp_2\ }{{x}, \many{z}}{H(\many{z}) \!\! }
  \arcr[2pt]
  \forall \many{z}, \hsll{\! H(\many{z})\! }{\iota:\ p'_1}{{x}}{Q({x}\many{z}) \!\! }
  \end{array}
  }
  {
  \hsllazy{\! P\ \ }{\iota : p_1; p'_1,~~S : \ \pp_2\ }{{x_1}, \many{x_2}}{Q({x_1}\many{x_2}) \!\! }
  }  
  }}
  \end{minipage}
}}
\caption{Selected structural \logic rules}
\label{log:struct}
\end{figure}

\begin{wrapfigure}[8]{r}{0.49\textwidth}
  \vspace{-15pt}
  \setlength{\abovecaptionskip}{7pt}
  \begin{minipage}{1\linewidth}
  \small
  \begin{prooftree}
    \AXC{\(\hslcc{P}{S_1:\pp_1}{H} \quad H\vdash\wp{S_2:\pp_2}{Q}\)}
    \RightLabel{\rulename{Cons}}
    \UIC{\(\hslcc{P}{S_1:\pp_1}{{\color{black} \wp{S_2:\pp_1}{Q}}}\)}
    \UIC{\(P \vdash \wp{S_1:\pp_1}{{\color{black} \wp{S_2:\pp_1}{Q}}}\)}
    \RightLabel{\rulename{WpNest}}
    \UIC{\(P \vdash \wp{S_1:\pp_1,\ S_2:\pp_2}{Q}\)}
    \UIC{\(\hslcc{P}{S_1:\pp_1,\ S_2:\pp_2}{Q}\)}
  \end{prooftree}
\end{minipage}
\caption{\rulename{SeqU} derivation using \rulename{WpNest}} 
\label{fig:deriv} 
\end{wrapfigure}
We start our overview of \logic reasoning principles by looking at its
structural rules (\autoref{log:struct}) that facilitate
transformations of the triples following the structure of the
\emph{index set}, thus enabling applications of the
program-driven \emph{lockstep} rules presented below.

For example, the rule \rulename{WpNest} exploits the nature of the
definition of \logic triples via the \textsf{wp} predicate to ``bring
forward'' a subset of the programs $\pp_1$ in the product (for a
particular subset $S_1$ of the index set), enabling more specialised
reasoning about this set in particular.
The utility of \rulename{WpNest} comes from its ability to derive
rules such as \rulename{SeqU} (as well as other structural rules from the overview).
\autoref{fig:deriv} shows this derivation by first replacing the
triple in the conclusion by its definition in terms of the weakest
preconditions, then applying \rulename{WpNest}, and then rewriting the
judgements in the \textsf{wp}-form back to the corresponding triples.
The full derivations of other rules from the overview section in the paper  are broadly similar, and are omitted here for brevity.

We have already seen an application of the rule \rulename{Product} in
the Section~2 of the paper.
The remaining structural rules from \autoref{log:struct} are standard
for separation logic-style proof systems. 
For example, \logic enjoys a familiar \rulename{Frame} rule that
naturally extends to hyper-heaps. 
%

\paragraph{Lockstep rules}
\label{sec:lockstepz}

\begin{figure}[t]
\setlength{\abovecaptionskip}{0pt}
\setlength{\belowcaptionskip}{-10pt}
{\small{
  \begin{subfigure}{0.48\textwidth}
    \begin{prooftree}
      \AXC{}
      \RightLabel{\tname{Ret}}
      \UIC{\(\hslc{\tt emp}{\kw{return}\ \many{v}}{\many{u}}{\pure{\many{v} = \many{u}}}\)}
    \end{prooftree}
  \end{subfigure}
  \begin{subfigure}{0.48\textwidth}
    \begin{prooftree}
      \AXC{}
      \RightLabel{\tname{Read}}
      \UIC{\(\hslc{\many{x}\mapsto\many{v}}{ !\many{x}}{\many{u}}{
        \many{x}\mapsto\many{v}\!\str\!\pure{\many{u} = \many{v}}}\)}
    \end{prooftree}
  \end{subfigure}
  \begin{subfigure}{0.3\textwidth}
    \vspace{2mm}
    \begin{prooftree}
      \AXC{}
      \RightLabel{\tname{Asn}}
      \UIC{\(\hslcc{\many{x}\mapsto\many{v}}{ \many{x}:=\many{u}}{
        \many{x}\mapsto\many{u}}\)}
    \end{prooftree}
  \end{subfigure}
  \begin{subfigure}{0.3\textwidth}
    \vspace{2mm}
    \begin{prooftree}
      \AXC{}
      \RightLabel{\tname{Fr}}
      \UIC{\(\hslcc{\many{x}\mapsto\many{v}}{ \kw{free}(\many{x})}{
        \tt emp}\)}
    \end{prooftree}
  \end{subfigure}
  \begin{subfigure}{0.32\textwidth}
    \vspace{2mm}
    \begin{prooftree}
      \AXC{}
      \RightLabel{\tname{Alc}}
      \UIC{\(\hslc{\tt emp}{\kw{alloc}(\many{v})}{\many{x}}{\many{x}\mapsto\many{v}}\)}
    \end{prooftree}
  \end{subfigure}
  \begin{subfigure}{0.5\textwidth}
    \vspace{2mm}
    \begin{prooftree}
      \AXC{}
      \RightLabel{\tname{Malloc}}
      \UIC{\(\hslc
        {\tt emp}
        {\kw{malloc}(\many{n})}
        {\many{x}}
        {\arrl{\many{x}}{0_{\many{n}}}}\)
        }
    \end{prooftree}
  \end{subfigure}
  \begin{subfigure}{0.4\textwidth}
    \vspace{2mm}
    \begin{prooftree}
      \AXC{}
      \RightLabel{\tname{MFree}}
      \UIC{\(\hslcc{\arrl{\many{x}}{\many{s}}}{\kw{mfree}(\many{x})}{
        \tt emp}\)}
    \end{prooftree}
  \end{subfigure}
  \begin{subfigure}{0.48\textwidth}
    \vspace{2mm}
    \begin{prooftree}
      \AXC{\(\hslcc{H}{\pp_1}{P}\quad\forall\many{v},\ \hslcc{P(\many{v})}{\pp_2[\many{v}/\many{x}]}{Q} \) }
      \RightLabel{\tname{Let}}
      \UIC{\(\hslcc{H}{\kw{let}\ \many{x}:= \pp_1\ \kw{in}\ \pp_2}{Q}\)}
    \end{prooftree}
  \end{subfigure}
  \begin{subfigure}{0.48\textwidth}
    \vspace{2mm}
    \begin{prooftree}
      \AXC{\(\hslcc{H}{\pp}{Q}\quad\forall i,
      \begin{array}{l} \many{v}(i) \Rightarrow \pp(i) = \pp_1(i) \\
                       \neg\many{v}(i) \Rightarrow \pp(i) = \pp_2(i)
         \end{array} \) }
      \RightLabel{\tname{If}}
      \UIC{\(\hslcc{H}{\kw{if}\ (\many{v})\ \{\pp_1\}\ \{\pp_2\} }{Q}\)}
    \end{prooftree}
  \end{subfigure}
\begin{subfigure}{0.5\textwidth}
  \vspace{2mm}
  \begin{prooftree}
    \AXC{}
    \RightLabel{\tname{Len}}
    \UIC{\(\hslc{\arrl{\many{x}}{\many{s}}}{\kw{length}(\many{x})}{\many{u}}{\pure{\forall i, \many{u}(i) = |\many{s}(i)|}}\)}
  \end{prooftree}
\end{subfigure}
\begin{subfigure}{0.48\textwidth}
  \vspace{2mm}
  \begin{prooftree}
    \AXC{\( \forall j\in S, \many{r}(j) = \many{s}(j)[\many{i}(j)\mapsto \many{v}(j)] \)}
    \RightLabel{\tname{AsnArr}}
    \UIC{\(\hslcc{\arrl{\many{x}}{\many{s}}}{ \many{x}[\many{i}]:=\many{v}}{
      \arrl{\many{x}}{\many{r}}}\)}
  \end{prooftree}
\end{subfigure}
  \begin{subfigure}{\textwidth}
    \vspace{2mm}
    \begin{prooftree}
      \AXC{\(\forall j \in S,\ 0 \leq \many{i}(j) < |\many{s}(j)|\)}
      \RightLabel{\tname{ReadArr}}
      \UIC{\(\hslc{\arrl{\many{x}}{\many{s}}}{\many{x}[\many{i}]}{\many{u}}{
        \arrl{\many{x}}{\many{s}}\!\str\!\pure{\forall j \in S, \many{u}(j) = \many{s}(j)(\many{i}(j)) }}\)}
    \end{prooftree}
  \end{subfigure}
}}
\caption{Lockstep rules}
\label{log:lockst}
\end{figure}

The core of any relational program logic is formed by so-called
\emph{lockstep} rules that enable reasoning about individual programs
in the product or about batches of programs that share the same
syntactic structure.
The lockstep rules of \logic are shown in \autoref{log:lockst}. 
In their presentation, we follow a few syntactic conventions.
First, by applying a program construction to a vector, we mean a
pointwise application: for example, for a hyper-value $\many{v}$ and
two product-programs, $\pp_1$ and $\pp_2$, indexed by a set $S$, for
any $i \in S$ the effect of programs is defined as follows:
\begin{equation*}
{\small{
(\kw{return}\ \many{v})(i) \eqdef \kw{return}\ \many{v}(i) \qquad (\kw{if}(\many{v})\{\pp_1\}\{\pp_2\})(i) \eqdef \kw{if}(\many{v}(i))\{\pp_1(i)\}\{\pp_2(i)\}
}}  
\end{equation*}
The same convention applies to hyper-heap assertions, \eg, points-to
assertions distribute similarly:
\begin{equation*}
{\small{
\many{x}\mapsto\many{v} \eqdef \bigstr_{i \in S} \many{x}(i) \mapsto
\many{v}(i)
}}  
\end{equation*}
Since lockstep rules work on syntactically same programs, we
omit the index set in the rules.

Most of the lockstep rules are just trivial extensions of
correspondent rules in a regular separation logic.
For example, \rulename{Ret} rule states that result of a program
product, each component of which is a \code{return}-statement
\code{return}~$\many{v}$, is a hyper-value $\many{u}$ is equal to the
one we return $\many{v}$.
The \rulename{Read} rule states that, if we read from a family of
pointers $\many{x}$, the result $\many{u}$ would be equal to a vector
constructed out of the contents of the pointers being read (\ie,
$\many{v}$), and that the hyper-heap will remain unchanged.
The \rulename{Asn}~rule says that, in order to write a value
$\many{u}$ to a vector of pointers $\many{x}$, one has to replace
their contents $\many{v}$ with $\many{u}$.
\rulename{Fr} states that after deallocation of a vector of pointes, the
resulting heap is empty.
\rulename{Alc} states that the results of a family of pointer allocations
are these pointers $\many{x}$, pointing to the assigned values $\many{v}$.
\rulename{Malloc, Mfree, Len, AsnArr} and \rulename{ReadArr}, are rules
for arrays extended to the hyper-case; we elaborate on the two of
them.
\rulename{Malloc}, ${{\kw{\small{malloc}}}}$ takes a length of an array
to be allocated. The result of the allocation is filled with zeros:
$0_{\many{n}}$ denotes a vector of sequences, each of which consists
of $\many{n}(i)$ zeros.
In the \rulename{AsnArr} rule, substitution
$\many{s}(j)[\many{i}(j)\mapsto\many{v}(j)]$ is defined only if
$0 \leq \many{i}(j) < |\many{s}(j)|$.
The rules \rulename{Let} and \rulename{If} are straightforward.


\end{document}
